\documentclass[preprint]{elsarticle}
\usepackage{amsmath, amssymb}
\usepackage[final]{epsfig}
\usepackage{color}
\usepackage{subfig, graphicx}
\usepackage{lineno}
\usepackage{suffix} 
\usepackage{bm}
\usepackage{hyperref}

\newcommand{\beq}{\begin{equation}}
\newcommand{\eeq}{\end{equation}}
\newcommand{\bdi}{\begin{displaymath}}
\newcommand{\edi}{\end{displaymath}}

\newcommand{\beqarr}{\begin{eqnarray}}
\newcommand{\eeqarr}{\end{eqnarray}}
\WithSuffix\newcommand\beqarr*{\begin{eqnarray*}}
\WithSuffix\newcommand\eeqarr*{\end{eqnarray*}}

\newcommand{\vs}{{\bf s}}
\newcommand{\vE}{{\bf E}}

\newcommand{\vx}{{\bf x}}

\newcommand{\Vex}{V_{\mathrm{ex}}}
\newcommand{\Veff}{V_{\mathrm{eff}}}
\newcommand{\Iind}{\ensuremath{I^{\mathrm{ind}}}}
\newcommand{\Iindpeak}{\ensuremath{I^{\mathrm{ind}}_{\mathrm{peak}}}}
\newcommand{\Vdpeak}{\ensuremath{V_{d,\mathrm{peak}}}}
\newcommand{\Iindmax}{\ensuremath{I^{\mathrm{ind}}_{\mathrm{max}}}}
\newcommand{\Idet}{\ensuremath{I_\mathrm{det}}}

\newcommand{\vEext}{\vE_{\mathrm{ext}}}
\newcommand{\vEsc}{\vE_{\mathrm{sc}}}
\newcommand{\Esc}{E_{\mathrm{sc}}^{x}}

\renewcommand{\vec}[1]{\ensuremath{\bm{\mathrm{#1}}}}
\newcommand{\ex}{\hat{\vec{x}}}

\newcommand{\Vbr}{V_{\mathrm{br}}}
\newcommand{\Ebr}{E_{\mathrm{br}}}
\newcommand{\Sbr}{K_{\mathrm{br}}}

\newcommand{\alphabr}{\alpha_{\mathrm{br}}}
\newcommand{\betabr}{\beta_{\mathrm{br}}}
\newcommand{\kappabarbr}{\bar\kappa_{\mathrm{br}}}
\newcommand{\Vsupply}{V_{\mathrm{supply}}}
\newcommand{\tmax}{t_{\mathrm{max}}}
\newcommand{\tdet}{t_{\mathrm{det}}}
\newcommand{\Vdet}{V_{\mathrm{det}}}

\newcommand{\deltaT}{\Delta t}
\newcommand{\tauQ}{\ensuremath{\tau_q}}
\newcommand{\FWHM}{\mathrm{FWHM}}
\DeclareMathOperator\artanh{artanh}
\newcommand{\avg}[1]{\left\langle #1 \right\rangle}
\newcommand{\eps}{\epsilon}

\newcommand{\micron}[1]{{#1}\,\text{\textmu{}m}}
\newcommand{\ps}[1]{{#1}\,\text{ps}}
\newcommand{\fF}[1]{{#1}\,\text{fF}}
\newcommand{\kohm}[1]{{#1}\,\text{k}\Omega}
\newcommand{\Vum}[1]{{#1}\,\text{V}/\text{\textmu{}m}}

\newcommand{\Nej}{N_e^j}
\newcommand{\Nhj}{N_h^j}
\newcommand{\Neehj}{N_{e\rightarrow eh}^j}
\newcommand{\Nhehj}{N_{h\rightarrow eh}^j}
\newcommand{\NeehjL}{N_{e\rightarrow eh}^{j\rightarrow j-1}}
\newcommand{\NeehjC}{N_{e\rightarrow eh}^{j\rightarrow j}}
\newcommand{\NeehjR}{N_{e\rightarrow eh}^{j\rightarrow j+1}}
\newcommand{\NhehjL}{N_{h\rightarrow eh}^{j\rightarrow j-1}}
\newcommand{\NhehjC}{N_{h\rightarrow eh}^{j\rightarrow j}}
\newcommand{\NhehjR}{N_{h\rightarrow eh}^{j\rightarrow j+1}}
\newcommand{\Po}{\mathrm{Po}}
\newcommand{\Mult}{\mathrm{Mult}}

\begin{document}


\begin{frontmatter}

\title{Passive quenching, signal shapes, and space charge effects\\ in SPADs and SiPMs}

\author[Oxford]{P. Windischhofer}
\author[CERN]{W. Riegler}

\address[Oxford]{University of Oxford}
\address[CERN]{CERN}

\begin{abstract}
\noindent
In this report we study the dynamics of passive quenching in a single-photon avalanche diode.
Our discussion is based on a microscopic description of the electron-hole avalanche coupled to the
equivalent circuit of the device, consisting of the quench resistor and the junction capacitance.
Analytic expressions for the resulting signal shape are derived from this model for simple electric
field configurations, and efficient numerical prescriptions are given for realistic 
device geometries.
Space charge effects arising from the avalanche are included using simulations.
They are shown to distort the signal shape, but alter neither its basic characteristics nor the 
underlying quenching mechanism.
\end{abstract}

\end{frontmatter}

\section{Introduction}
\label{sec:introduction}
\noindent
Single-photon avalanche diodes (SPADs) and silicon photomultipliers (SiPMs) are Geiger-mode avalanche detectors with
direct single-particle sensitivity to photons and charged particles \cite{renker:2006, acerbi:2019, spad2}.
In these devices, the small amount of charge deposited by the primary particle triggers a self-sustaining 
electron-hole avalanche which leads to the creation of a detectable electric signal.
The avalanche develops in a thin region with very high electric fields, created in a p-n diode which is
initially reverse-biased above its breakdown voltage.

The resulting exponential growth of the avalanche must be stopped (``quenched'') at a suitable moment so that all
charge carriers are removed from the junction and the detector can register the next event.
In passively quenched devices, the diode is decoupled from the bias supply through a series resistance.
The charge produced by the avalanche then reduces the electric field in the junction and suppresses the avalanche
growth.
Quenching occurs when the voltage across the diode drops below the breakdown limit.
Once all charges have drifted out of the junction, the original bias voltage is restored.

The resulting voltage and current waveforms form the output signal.
The signal properties are ultimately determined by the complicated interplay of impact ionisation, 
the quenching circuit, and the diode capacitance.
The operating conditions are typically chosen such that the achieved signal charge is around $10^4$--$10^5$ electrons, 
i.e.~very large avalanches must be considered.

Existing analyses of the signal formation in passively quenched SPADs either model the avalanche discharge and quenching
purely in terms of an equivalent electrical circuit \cite{seifert:2009, cova:1996, haitz:1964}, or use a coarse
effective description of the avalanche development, e.g.~in terms of rate equations \cite{inoue:2020, inoue:2021}.
In this report, we construct a simple analytic quenching model which is directly based on the microscopic dynamics of electron-hole
avalanches.
It naturally connects to our previous treatment of avalanche statistics given in Ref.~\cite{spad1}.
As a result, our model describes the detector signal in terms of the underlying material parameters and the device 
geometry.
It can therefore guide the evaluation of specific detector designs and allows direct comparisons to be made with measurements.

This paper is structured as follows.
Section \ref{sec:avalanche_properties} briefly reviews the avalanche formation and summarises properties 
of large electron-hole avalanches used to formulate the quenching model.
Analytic predictions for the signal shape and the spatial distributions of charge carriers in the high-field region
are derived in Section \ref{sec:passive_quenching}.
The impact of the space charge field of the avalanche on the quenching process is then studied in 
Section \ref{sec:space_charge}. 

\section{Properties of large electron-hole avalanches}
\label{sec:avalanche_properties}
\noindent
The development of the electron-hole avalanche is determined by the transport of charge carriers of both polarities
through the multiplication region and their interactions with the material.
For sufficiently strong electric fields, both electrons and holes can cause impact ionisation and produce additional
electron-hole pairs.
The probability for a charge carrier to undergo a multiplication reaction may be parameterised in terms of the material
parameters and, in general, the past history of the carrier.
The drifting carriers in the avalanche induce a current $\Iind$ on the metallised device contacts, which
may be computed with the Ramo-Shockley theorem \cite{ramo:1939, shockley:1938}.

Impact ionisation is a stochastic process. 
The avalanche development is thus subject to fluctuations, which are particularly pronounced at early times when the total 
number of participating charge carriers is still small.
For avalanches containing many charge carriers, these fluctuations average out and the further evolution appears deterministic.
(The continued evolution of an avalanche containing $N$ charges will generate relative fluctuations with a magnitude proportional to
$1/\sqrt{N}$, as discussed in Section 3.5 of Ref.~\cite{spad1}.)
It is useful to introduce a current scale $\Idet$ to separate these two regimes, chosen such that avalanche fluctuations
are important whenever $\Iind \ll \Idet$ and a deterministic description of the avalanche is appropriate when $\Iind \gtrsim \Idet$.
The current $\Idet$ therefore \textsl{defines} the level of fluctuations that are deemed ``important'' in a particular situation.
It implements a purely semantic definition, and expressions for physical observables do not depend explicitly on $\Idet$.

\begin{figure}[tp]
  \centering
  \includegraphics[width=7cm]{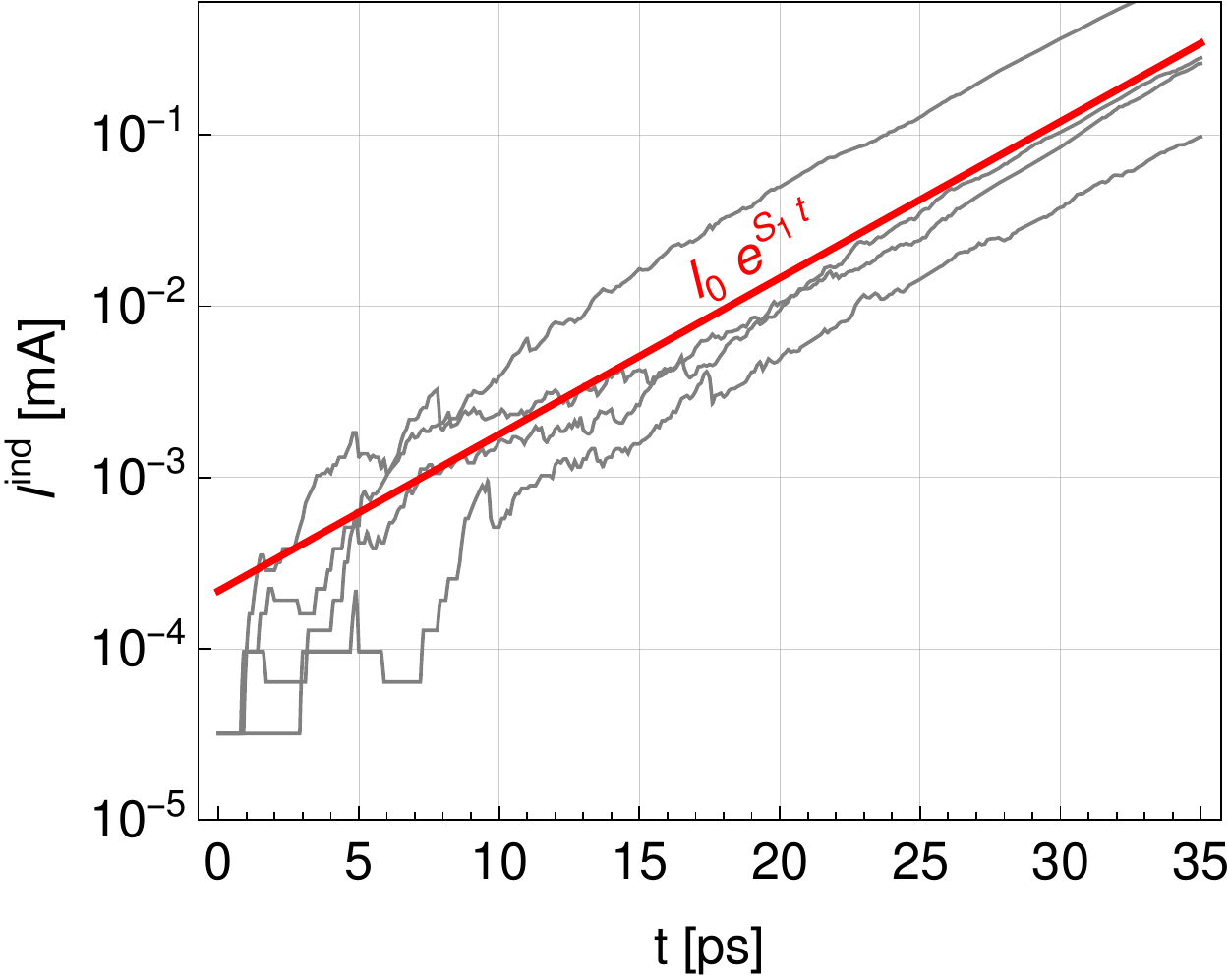}
  \caption{(Colour online.) The grey lines represent typical simulated avalanche events, each starting
  from a single primary electron. 
  For large avalanches, the evolution of $\Iind$ in each event becomes proportional 
  to the average current $\avg{\Iind}$, parameterised by $I_0$ and $S_1$, as indicated by the red line.}
  \label{fig:avalanche_samples_early}
\end{figure}

Our model of the quenching process is based on the following two general properties of large electron-hole avalanches 
($\Iind \gtrsim \Idet$) developing in a time-independent electric field, illustrated in Fig.~\ref{fig:avalanche_samples_early}:
\begin{enumerate}[(i)]
    \item The shape of the induced current $\Iind$ is identical for each avalanche event.
    It is directly proportional to the average current $\avg{\Iind}$, where the average is taken over all avalanche evolutions
    starting from the same initial conditions,
    \beq
    \Iind(t) = k \avg{\Iind(t)}.
    \label{eq:large_avalanche_fluctuations}
    \eeq
    The proportionality factor $k$ is a random variable.
    The resulting fluctuations around the average evolution determine the time resolution, but are irrelevant for the study of the quenching process.
    \item The average current $\avg{\Iind}$ evolves exponentially as 
    \beq
    \avg{\Iind(t)} = I_0 \exp{(S_1 t)},
    \label{eq:av_current}
    \eeq
    and is parameterised by the dimensionful quantities $I_0$ and $S_1$.
    The current $I_0$ depends on the number and the spatial distribution of the primary charges initiating the avalanche.
    The parameter $S_1$ represents the asymptotic net growth rate of the avalanche, i.e.~it takes into account the generation of charges in
    the high-field region and their outflow across its boundaries.
    It depends on the device geometry and the reverse bias voltage $V_d$ applied across the junction, i.e.~$S_1 = S_1(V_d)$.
    In the following, the breakdown voltage $\Vbr$ is defined such that $S_1(\Vbr) = 0$. 
    This implies that $S_1(V_d) > 0$ for $V_d > \Vbr$, i.e.~exponentially growing avalanches are possible if the
    device is operated ``above breakdown''.
    For a bias voltage below the breakdown limit, $V_d < \Vbr$, impact ionisation cannot sustain the discharge.
    The avalanche current then decays exponentially, i.e.~$S_1(V_d) < 0$.
\end{enumerate}
For a given device and electric field configuration, the parameters $S_1$ and $I_0$ are always accessible through Monte Carlo (MC)
simulations of the avalanche development (such as the one in Fig.~\ref{fig:avalanche_samples_early}); efficient numerical prescriptions 
for quasi-one-dimensional situations are referenced below.
The above properties arise as explicit predictions of the avalanche model studied analytically in a one-dimensional geometry 
in Ref.~\cite{spad1}. It is briefly reviewed in the following.
We use this model to illustrate our results in Sections \ref{sec:passive_quenching} and \ref{sec:space_charge}.
However, our description of the quenching process is more general and applies to all scenarios in which the avalanche 
obeys the properties (i)--(ii) above, including realistic three-dimensional device geometries.

\paragraph{One-dimensional memoryless avalanche model}

\begin{figure}[tp]
  \centering
  \subfloat[]{\raisebox{0.02\textwidth}{\includegraphics[height=0.25\textwidth]{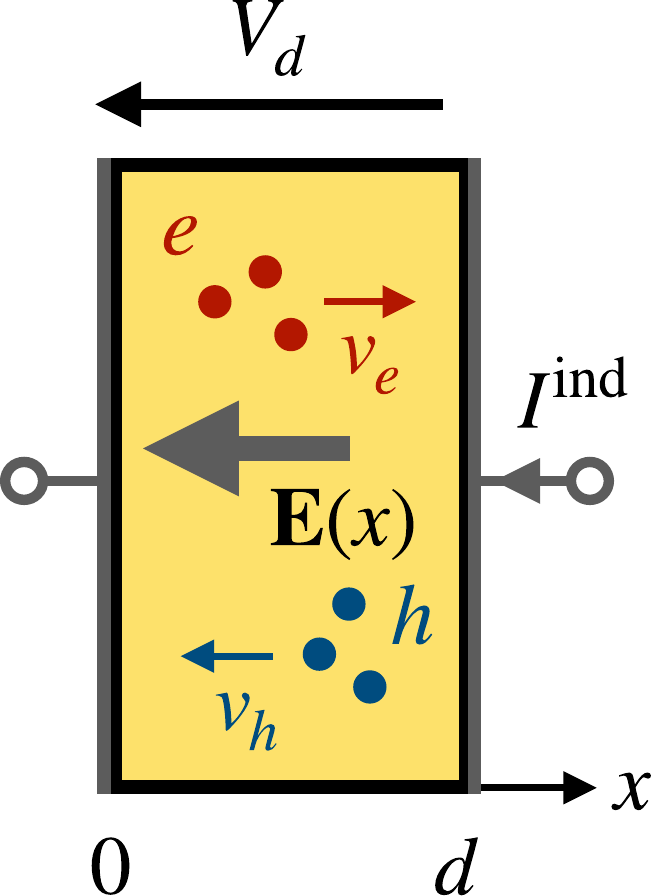}}\label{subfig:bounded_avalanche_region}}\qquad\quad
  \subfloat[]{\raisebox{0.02\textwidth}{\includegraphics[height=0.25\textwidth]{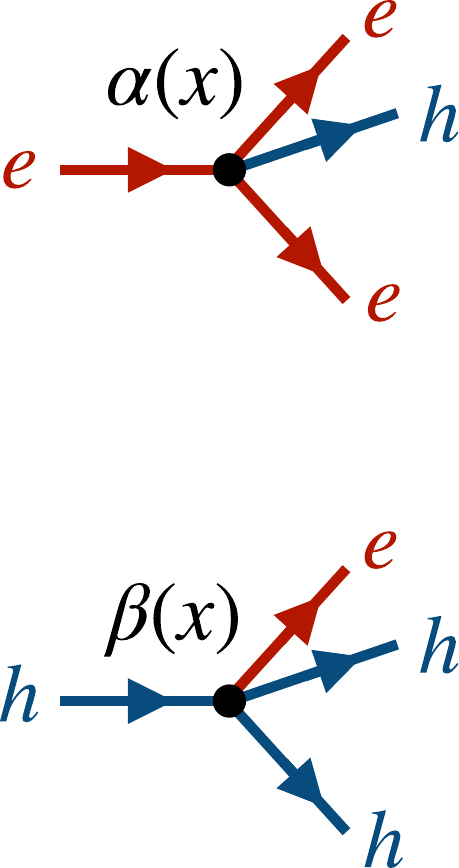}}\label{subfig:impact_ionisation}}\qquad
  \subfloat[]{\raisebox{0.025\textwidth}{\includegraphics[height=0.25\textwidth]{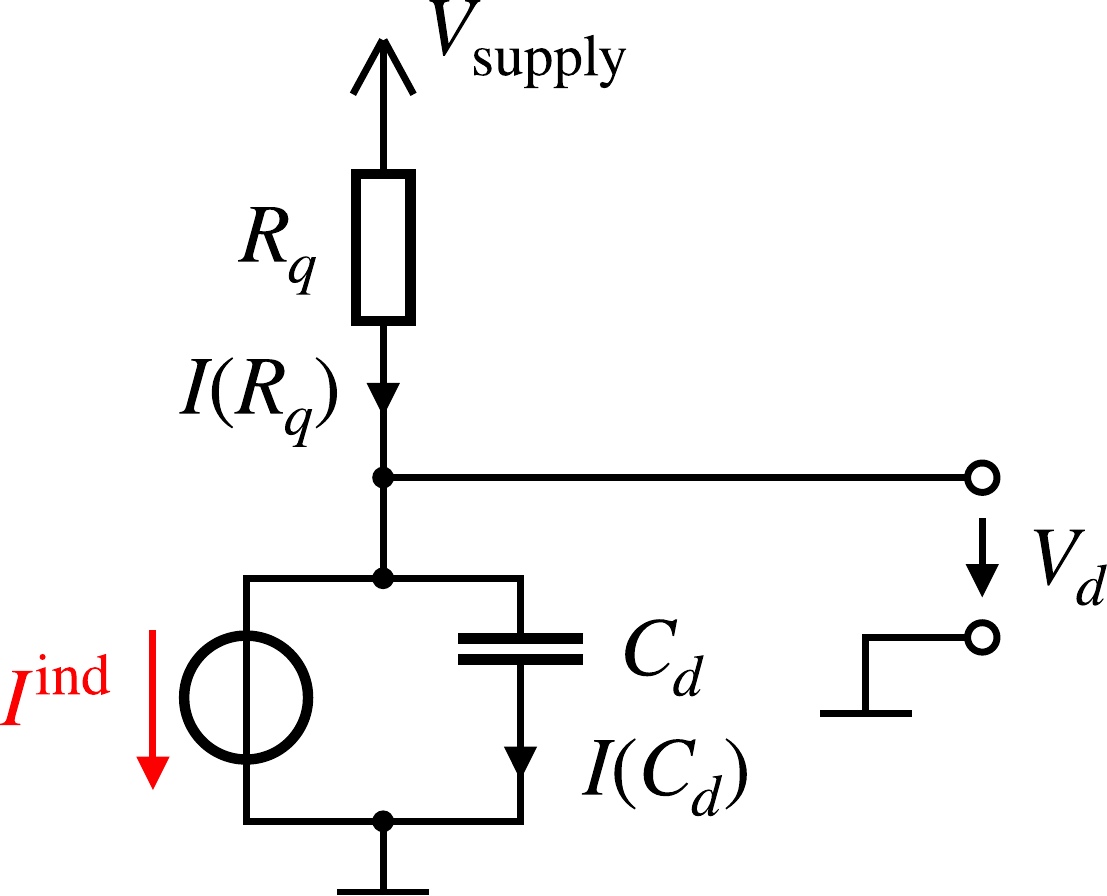}\label{subfig:quenching_circuit}}}
  \caption{
    \protect\subref{subfig:bounded_avalanche_region} Electrons and holes drift in opposite directions with velocities 
    $v_e$ and $v_h$ in the electric field $\vE(x)$ of the p-n junction, inducing a current $\Iind$.
    \protect\subref{subfig:impact_ionisation} Both charge carrier species can undergo impact ionisation and produce 
    additional electron-hole pairs, parameterised by the impact ionisation coefficients $\alpha$ and $\beta$.
    \protect\subref{subfig:quenching_circuit} Equivalent circuit used for the analysis of the quenching process,
    consisting of the diode capacitance $C_d$ and the quench resistor $R_q$. 
    The current source represents the induced current $\Iind$.
  }
  \label{fig:quenching_dynamics_circuit}
\end{figure}

The diameter $D$ of a typical SPAD pixel is significantly larger than the longitudinal thickness $d$ of the high-field 
region, typically $D / d \gtrsim 10$.
Close to the centre of a pixel, this results in a quasi-one-dimensional field geometry through which electrons and holes drift in opposite directions 
with velocities $v_e$ and $v_h$, respectively.
The drift direction is taken to be along the $x$-axis (Fig.~\ref{subfig:bounded_avalanche_region}).

The one-dimensional model of Ref.~\cite{spad1} describes the avalanche development in this geometry.
In this model, impact ionisation is parameterised by the impact ionisation coefficients $\alpha$ and $\beta$,
which represent the probability per unit length for an electron or hole to undergo a multiplication reaction
(Fig.~\ref{subfig:impact_ionisation}).
This probability is assumed to depend only on the local electric field strength $\vE(x)$, i.e.~$\alpha(x) = \alpha(\vE(x))$
and $\beta(x) = \beta(\vE(x))$.
Any dependence on the history of the charge carriers is neglected, i.e.~the avalanche is assumed to be a memoryless
stochastic process.

For general field profiles $\vE(x)$, the growth parameter $S_1$ may be obtained from the numerical solution of the
differential equations describing the evolution of the average charge content of the avalanche, Eqs.~30 and 31 in Ref.~\cite{spad1}.
These equations also determine the spatial carrier densities in the junction.
As shown in Appendix B of Ref.~\cite{spad2}, the breakdown condition $S_1 = 0$ may be expressed in terms of the
impact ionisation coefficients as
\beq
\int_0^d dx \, \alpha(x) \exp\left[-\int_0^x dx' \, \left(\alpha(x') - \beta(x')\right)\right] = 1.
\label{eq:breakdown_condition}
\eeq
It is thus equivalent to the breakdown condition of Ref.~\cite{stillman:1977},
which evaluates the point at which the gain for a constant current injected into the junction diverges.

For constant fields, $S_1$ may be expressed as the product of the effective carrier drift velocity 
$v^* = 2 v_e v_h / (v_e + v_h)$ and the growth factor $\gamma_1$, i.e.~$S_1 = \gamma_1 v^*$.
The latter is defined in Eq.~35 in Ref.~\cite{spad1}.
The drift velocities for electrons and holes, $v_e$ and $v_h$, are themselves dependent on the local electric field.
The spatial carrier densities for electrons and holes, $n_e(x, t)$ and $n_h(x, t)$, are proportional to the functions $f_{\lambda_1}^e(x, t)$ and $f_{\lambda_1}^h(x, t)$, respectively.
Their analytic expressions are given in Eqs.~36 and 37 of Ref.~\cite{spad1}.
The numerical parameter $\lambda_1$ is related to the time constant $S_1$ and defined in Eq.~38 of Ref.~\cite{spad1}.

The proportionality factor $k$ in Eq.~\ref{eq:large_avalanche_fluctuations} is approximately distributed according to a 
gamma distribution with shape parameter $A$ and scale parameter $A^{-1}$ (cf.~Section 3.6 of Ref.~\cite{spad1}),
\beq
p(k) dk = \frac{k^{A-1} e^{-k/A^{-1}}}{\Gamma(A)A^{-A}} dk,
\label{eq:k_distrib}
\eeq
where $\Gamma(z)$ is the gamma function, and the avalanche parameter $A$ is defined as
\beq
A = \frac{\alpha v_e N_e^0 + \beta v_h N_h^0}{\alpha v_e + \beta v_h}
\label{eq:avpar_def}
\eeq
for position-independent electric fields.
In Eq.~\ref{eq:avpar_def}, $N_e^0$ and $N_h^0$ label the numbers of primary electrons and holes, respectively.
For general field profiles $\vE(x)$, the avalanche parameter may be computed as shown in Eq.~79 of Ref.~\cite{spad1}.

\section{Passive quenching in the absence of space charge effects}
\label{sec:passive_quenching}
\noindent
To describe the quenching process, the avalanche evolution summarised in Section \ref{sec:avalanche_properties} 
is coupled to the electrical equivalent circuit of a single SPAD (Fig.~\ref{subfig:quenching_circuit}).
In this circuit, the p-n junction is represented by its capacitance $C_d$, and the current $\Iind$ acts as
a parallel current source.
(Note that the dynamic resistance $R_d$ of the diode is automatically included through the dependence of the carrier 
drift velocities $v_e$ and $v_h$ on the electric field and is thus not present as an explicit circuit component.)
In the absence of free charge carriers in the junction, $\Iind = 0$ and $V_d = \Vsupply$.
The excess voltage $\Vex$ is defined as $\Vex = \Vsupply - \Vbr > 0$, where the breakdown voltage $\Vbr$ is determined
by Eq.~\ref{eq:breakdown_condition}.

The time-dependent voltage $V_d(t)$ determines the electric field  $\vE(x,t) = \vE(x, V_d(t))$ which is relevant for the development of the avalanche.
To ensure quenching, the decoupling resistor $R_q$ must be chosen such that the recharging time constant $R_q C_d$ 
is much longer than the time scale $1 / S_1(\Vsupply)$ on which the avalanche develops.
The initial phase of the quenching process thus remains unchanged as $R_q\rightarrow\infty$.

In this dynamical situation, the instantaneous growth rate of the avalanche, $S_1(t)$, becomes itself time-dependent.
Provided that $V_d(t)$ changes on time scales that are long compared to the internal time constants of the avalanche, 
the changing growth rate may be approximated as $S_1(t) \approx S_1(V_d(t))$, where $S_1(V_d)$ is the function introduced in 
Section \ref{sec:avalanche_properties} for \textsl{static} situations.
This ``adiabatic approximation'' is shown below to be applicable to practically relevant situations.
For $R_q\rightarrow \infty$ and $\Iind \gtrsim \Idet$, the combined deterministic system of avalanche and equivalent circuit is 
then described by the equations
\beqarr
\frac{d \Iind}{dt} &=& S_1(V_d) \Iind,\label{eq:Iind_evolution}\\
\frac{d V_d}{dt} &=& -\frac{\Iind}{C_d}.\label{eq:Vd_evolution}
\eeqarr
More complicated equivalent circuits that include additional parasitic elements or finite values of the
quench resistance $R_q$ are readily accommodated using the same steps.
Eq.~\ref{eq:Iind_evolution}, which describes the evolution of the avalanche, must then be solved together with
the corresponding circuit equations.
For the situation discussed here, Eqs.~\ref{eq:Iind_evolution}--\ref{eq:Vd_evolution} may be solved numerically for
general $S_1(V_d)$.
This is similar to the rate model constructed in Ref.~\cite{inoue:2020}, but in contrast to the calculations
performed there automatically includes boundary effects.

In many practically relevant situations, $\Vex / \Vbr \ll 1$, i.e.~the voltage $V_d$ remains close to $\Vbr$ during the full
evolution of the system.
In this case, the function $S_1(V_d)$ in Eq.~\ref{eq:Iind_evolution} may be replaced by its linear expansion around
$\Vbr$.
With the breakdown condition $S_1(\Vbr) = 0$, this gives
\beq
S_1(V_d) \approx \frac{d S_1}{d V_d}\bigg\rvert_{\Vbr} \cdot (V_d - \Vbr) =: \Sbr \cdot (V_d - \Vbr).
\label{eq:linear_approximation_S1}
\eeq
The parameter $\Sbr$ may be easily obtained through finite differences provided that $S_1(V_d)$ is known
around $V_d \approx \Vbr$.
An analytic expression for $\Sbr$ for a memoryless avalanche and position-independent fields is given in 
Eq.~\ref{eq:model_parameter} below.

If the primary charge deposit occurs at $t=0$, Eqs.~\ref{eq:Iind_evolution}--\ref{eq:Vd_evolution} must be solved for 
$t \geq \tdet$, where $\tdet$ is the time at which the avalanche current first exceeds $\Idet$.
The time $\tdet$ is a random variable whose statistics are further discussed below.
Defining $V_d(\tdet) = \Vdet$, the relevant initial condition is $\Iind(\tdet) = \Idet$, leading to the following
analytic solutions,
\beqarr
\Iind(t) &=& \frac{C_d \Veff}{\tauQ} \left[1 -
\tanh^2\left(\frac{t-\tdet-\deltaT}{\tauQ}\right)\right],\label{eq:Iind_adiabatic_solution}\\
V_d(t) &=& \Vdet - \Veff\left[\tanh\left(\frac{t-\tdet-\deltaT}{\tauQ}\right) +
\tanh\left(\frac{\deltaT}{\tauQ}\right)\right],\label{eq:Vd_adiabatic_solution}
\eeqarr
with
\beqarr
\Veff = (\Vdet-\Vbr) \sqrt{1 + \frac{2 \Idet}{C_d \Sbr (\Vdet-\Vbr)^2}},
\label{eq:Veff_def}
\eeqarr
the quenching time constant $\tauQ$,
\beq
\tauQ = \frac{2}{\Sbr \Veff},
\label{eq:tauQ_def}
\eeq
and the time $\deltaT$,
\beq
\deltaT = \tauQ \artanh\left(\frac{\Vdet - \Vbr}{\Veff}\right).
\label{eq:deltaT_def}
\eeq
The quantity $\tauQ$ parameterises the time scale on which $V_d$ changes as the avalanche is quenched.
For Eqs.~\ref{eq:Iind_adiabatic_solution}--\ref{eq:Vd_adiabatic_solution} to be a valid description of the quenching process,
i.e.~for the adiabatic approximation to be applicable, $\tauQ$ must be larger than the transit time $d/v^*$.
The latter approximates the time scale on which the spatial charge distribution in the junction, and therefore also the
instantaneous growth rate $S_1(t)$, react to changes of $V_d$.

The general expressions in Eqs.~\ref{eq:Iind_adiabatic_solution}--\ref{eq:Vd_adiabatic_solution} can be further simplified.
For practical devices, the diode capacitance $C_d$ typically amounts to a few femtofarads.
With the elementary charge $e_0$, around $C_d \Vex / e_0 \sim 10^4$ charge carriers need to be produced by the avalanche
in the high-field region before any significant discharge of $C_d$ occurs, i.e.~any avalanche fluctuations remaining during
the quenching process amount to a few percent at most.
Working to this accuracy, $V_d(\tdet) = \Vdet \approx \Vsupply$ and $\Vdet - \Vbr \approx \Vex$; furthermore, $\Idet$ can be 
selected so that $\Idet \ll C_d \Sbr (\Vdet - \Vbr)^2$.
In this limit, Eq.~\ref{eq:Veff_def} simplifies to $\Veff \approx \Vex$ and the quenching time constant is directly given in terms of 
the excess voltage,
\beq
\tauQ \approx \frac{2}{\Sbr \Vex},
\label{eq:tauQ_def_approx}
\eeq
and the time $\Delta t$ defined in Eq.~\ref{eq:deltaT_def} becomes
\beq
\deltaT \approx \frac{\tauQ}{2} \log \frac{2 C_d \Sbr \Vex^2}{\Idet} \approx 
\frac{1}{S_1(\Vsupply)} \log \frac{2 C_d \Sbr \Vex^2}{\Idet}.
\label{eq:deltaT_def_approx}
\eeq
With this, $\deltaT / \tauQ \gg 1$ and
Eqs.~\ref{eq:Iind_adiabatic_solution}--\ref{eq:Vd_adiabatic_solution} simplify to
\beqarr
\Iind(t) &=& \frac{C_d \Vex}{\tauQ} \left[1 - \tanh^2\left(\frac{t-\tdet -  \deltaT}{\tauQ}\right)\right],\label{eq:Iind_adiabatic_solution_approx}\\
V_d(t) &=& \Vsupply - \Vex\left[1 + \tanh\left(\frac{t-\tdet- \deltaT}{\tauQ}\right)\right],\label{eq:Vd_adiabatic_solution_approx}
\label{eq:adiabatic_solution}
\eeqarr
and the adiabatic approximation is valid provided that $\Sbr \Vex \lesssim 2 v^* / d$.
The current signal $\Iind$ attains a peak value of $\Iindmax = C_d \Vex / \tauQ$ at time $\tmax = \tdet + \deltaT$.
If $t=0$ is instead regarded as the position of the peak, the signal shape reads
\beqarr
\Iindpeak(t) &=& \frac{C_d \Vex}{\tauQ} \left[1 - \tanh^2\left(\frac{t}{\tauQ}\right)\right],\\
\Vdpeak(t) &=& \Vsupply - \Vex\left[1 + \tanh\left(\frac{t}{\tauQ}\right)\right].
\eeqarr
Its evolution is symmetric around this peak with a FWHM of
\beq
\FWHM\left(\Iind\right) = 2\, \tauQ \artanh \frac{1}{\sqrt{2}} \approx 1.76\, \tauQ.
\eeq
The observable voltage step $\Delta V_d = \Vsupply - \lim_{t\rightarrow\infty} V_d(t)$
is 
\beq
\Delta V_d = 2 \Vex,
\label{eq:delta_Vd}
\eeq
and $V_d$ falls from $\Vsupply - 0.1 \Delta V_d$ to $\Vsupply - 0.9 \Delta V_d$ within 
about $2.2\,\tauQ$.
The total signal charge is 
\beq
Q = \int_{-\infty}^{\infty} dt\, \Iind(t) = 2 C_d \Vex.
\label{eq:signal_Q}
\eeq
The signal charge is thus directly proportional to the excess voltage at which the junction is operated and
does not depend on the primary charge deposit which initiated the avalanche.
The physical origin of the result in Eq.~\ref{eq:delta_Vd} is easily understood.
As long as $V_d(t) > \Vbr$, impact ionisation is self-sustaining and the avalanche grows exponentially.
The avalanche, and the induced current $\Iind$, consequently attains its maximum size when $V_d(\tmax) = \Vbr = \Vsupply - \Vex$.
Impact ionisation continues to occur for $V_d(t) < \Vbr$, but is no longer self-sustaining.
The current $\Iind$ then decays exponentially, reducing $V_d$ below $\Vbr$ and increasing the output voltage
step $\Delta V_d$ beyond $\Vex$.
In situations where the approximation in Eq.~\ref{eq:linear_approximation_S1} is appropriate, the collapse of the
avalanche exactly mirrors its initial growth; the evolution of $V_d(t)$ is then symmetric around $\Vbr$ and
so $\Delta V_d = G\, \Vex$ with $G = 2$.
Note that this conclusion deviates from other commonly-used quenching models, which take 
the avalanche discharge to stop already at $V_d(t) \approx \Vbr$ and thus lead to $G \approx 1$ 
\cite{seifert:2009, cova:1996, haitz:1964}.
These differences deserve a detailed comparison and discussion, which we carry out at the end of this Section.

The timing of the detector signal also carries important information.
The average time of maximum current,
\beq
\avg{\tmax} = \avg{\tdet} + \deltaT,
\eeq
is experimentally accessible in a setup where the primary charge deposit occurs at a well-defined time, e.g.~where charges are 
deposited by means of a laser pulse or a charged-particle beam.
It contains information about the early evolution of the avalanche (through $\avg{\tdet}$), as well as the quenching process itself
(through $\deltaT$).
Eqs.~\ref{eq:large_avalanche_fluctuations} and \ref{eq:av_current} result in the following expression for $\avg{\tdet}$,
\beq
\avg{\tdet} = \frac{1}{S_1(\Vsupply)}\left(\log\frac{\Idet}{I_0} - \avg{\log k}\right),
\eeq
which, together with Eq.~\ref{eq:deltaT_def_approx}, gives for $\avg{\tmax}$
\beq
\avg{\tmax} = \frac{1}{S_1(\Vsupply)}\left(\log\frac{2 C_d \Sbr \Vex^2}{I_0} - \avg{\log k}\right).
\label{eq:tmax}
\eeq
The logarithmic expectation $\avg{\log k}$ depends on the magnitude of the avalanche fluctuations;
an analytic expression valid for the memoryless avalanche model is derived in Eq.~\ref{eq:tmax_memoryless} below.

\paragraph{Quenching for one-dimensional memoryless avalanche}

\begin{figure}[tp]
 \begin{center}
  \subfloat[]{\includegraphics[width=7cm]{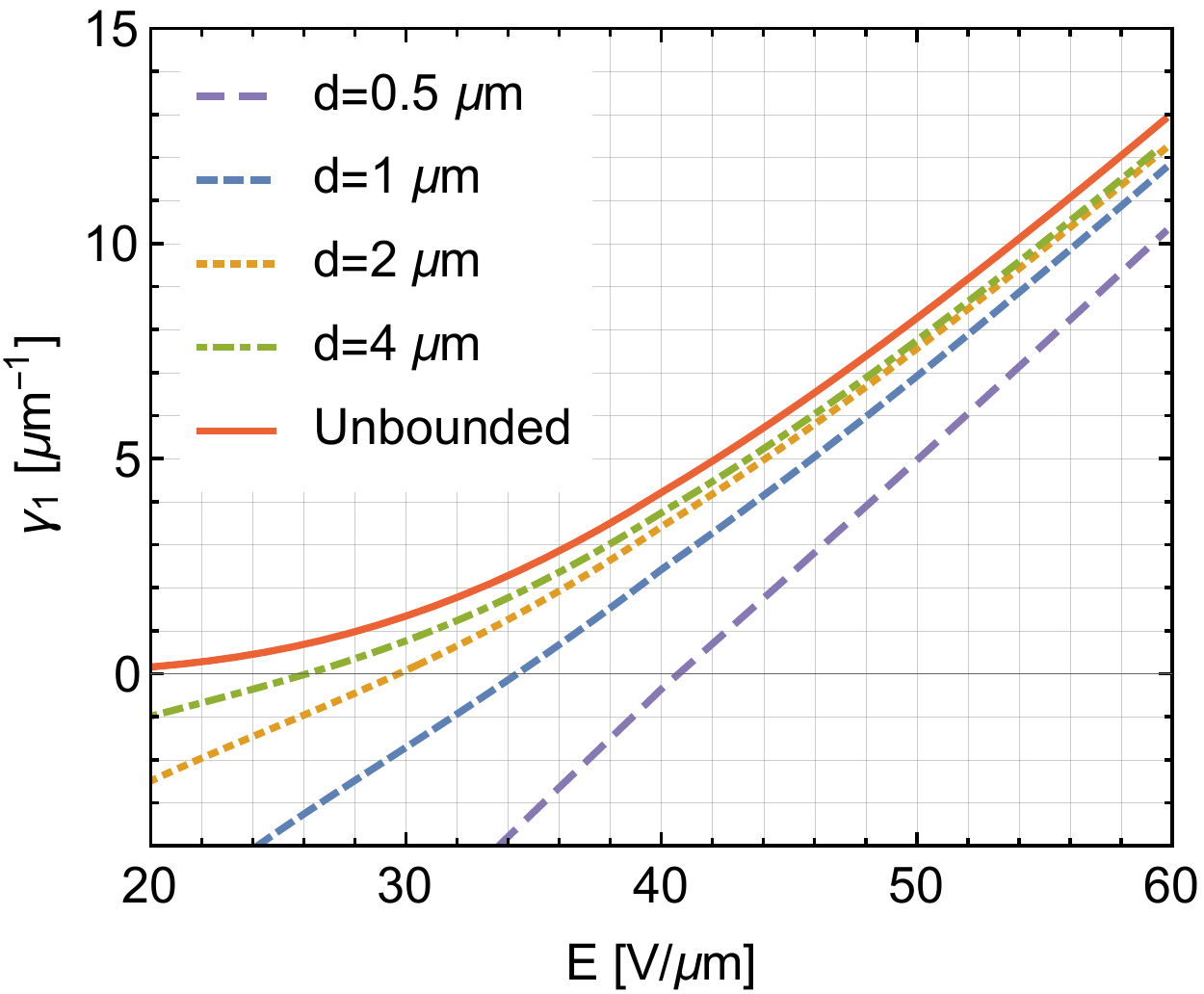}\label{subfig:gamma_field}}\qquad
  \subfloat[]{\includegraphics[width=7cm]{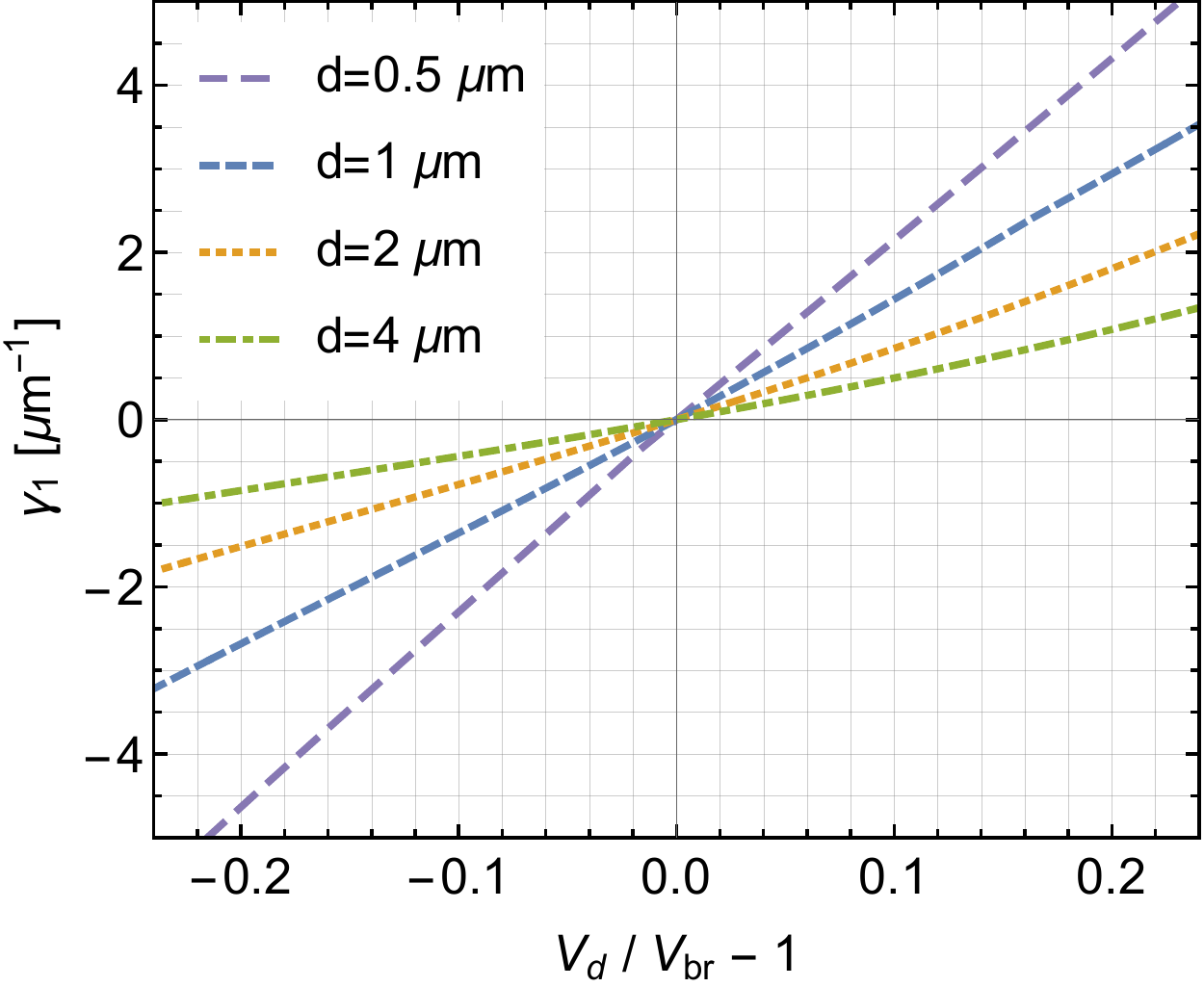}\label{subfig:gamma_voltage}}
   \caption{
    \protect\subref{subfig:gamma_field} Dependence of the growth parameter $\gamma_1$ on the electric field and the thickness $d$,
    for the case of silicon and the impact ionisation coefficients of Ref.~\cite{overstraeten:1969}.
    The limit of an unbounded multiplication region ($d\rightarrow\infty$) is also shown.
    \protect\subref{subfig:gamma_voltage} The parameter $\gamma_1$ is expressed in terms of the voltage $V_d$ close to
    the breakdown point.
   }
  \label{fig:gamma}
  \end{center}
\end{figure}

We illustrate these results for the case of a simplified SPAD, for which we approximate the field distribution in the 
multiplication region with a position-independent, one-dimensional electric field.
The field is thus taken to be $\vE(x, V_d) = -\ex \, V_d / d$, where $\ex$ is the unit vector along $x$ 
(cf.~Fig.~\ref{subfig:bounded_avalanche_region}).
In this case, the breakdown voltage is related to the breakdown field $\Ebr$ as $\Vbr = \Ebr d$.
The effective drift velocity $v^*$ typically saturates in the high-field region, i.e.~becomes independent of $V_d$.
Using the memoryless avalanche model from Section \ref{sec:avalanche_properties}, $S_1(V_d) = \gamma_1(V_d) v^*$ and
\beq
\Sbr = \frac{v^*}{d}\frac{d \gamma_1}{d E}\bigg\rvert_{\Ebr} =
  \frac{\left(\alphabr' + \betabr'\right)v^*}{2 d} + 
  \frac{\left(\sinh 2\kappabarbr - 2\kappabarbr \right) \left(\alphabr'\betabr + \alphabr \betabr' \right) v^*}
       {2\kappabarbr \left(\cosh 2\kappabarbr - 1 \right) - d \left(\sinh 2\kappabarbr - 2\kappabarbr \right) (\alphabr + \betabr)},
\label{eq:model_parameter}
\eeq
where $\alphabr = \alpha(\Ebr)$, $\betabr = \beta(\Ebr)$, $\alphabr' = d\alpha/dE\rvert_{\Ebr}$, $\betabr' = d\beta/dE\rvert_{\Ebr}$, and
$\kappabarbr = d |\alphabr - \betabr| / 2$.
In the adiabatic approximation, predictions for the charge carrier densities of electrons and holes are proportional to the functions
$f^e_{\lambda_1(V_d)}$ and $f^h_{\lambda_1(V_d)}$.
The linear expansion in Eq.~\ref{eq:linear_approximation_S1} describes $\gamma_1(V_d)$ to excellent precision for typical 
values of $\Vex$.
This is visualised in Fig.~\ref{fig:gamma} for silicon, using the parameterisations of the impact ionisation coefficients from
Ref.~\cite{overstraeten:1969}.

With the distribution of $k$ in Eq.~\ref{eq:k_distrib}, $\avg{\log k} = \psi(A) - \log(A)$ with the digamma function $\psi(z)$,
and the expression for $\avg{\tmax}$ in Eq.~\ref{eq:tmax} becomes
\beq
\avg{\tmax} = \frac{1}{S_1(\Vsupply)}\left(\log\frac{2 C_d \Sbr \Vex^2}{I_0} + \log(A) - \psi(A)\right),
\label{eq:tmax_memoryless}
\eeq
and $I_0$ may be computed for a given initial charge deposit as shown in the paragraph following Eq.~46 
in Ref.~\cite{spad1}.

The analytic results derived above are compared to a MC implementation of the memoryless avalanche model, 
coupled to the quenching circuit of Fig.~\ref{subfig:quenching_circuit}.
The avalanche simulation is performed as described in the Appendix, using the impact ionisation coefficients 
for silicon from Ref.~\cite{overstraeten:1969} and saturated drift velocities $v_e=v_h=v^*=\micron{0.1}/\mathrm{ps}$ \cite{canali:1975}.
The circuit equations are integrated using the forward Euler method.
The induced current $\Iind$ is computed assuming the weighting field of a parallel-plate geometry with
an electrode spacing of $d$ for the multiplication region.
A dedicated conversion layer, common in SPADs optimised for infrared photon detection, is not included.
The full simulation is referred to as ``memoryless avalanche MC + circuit'' (MAMC + circuit).
For the comparisons shown here, the thickness $d$ of the high-field region is chosen as $d = \micron{0.5}$ and its diameter is $D = \micron{10}$.
Assuming a parallel field geometry and using $\eps_r = 11.7$ \cite{dunlap:1953}, the junction capacitance is $C_d\approx\fF{16}$.
A quench resistor with $R_q=\kohm{200}$ is used.
The breakdown voltage for this geometry is $\Vbr \approx 20.34$\,V
and with Eq.~\ref{eq:model_parameter} the model parameter $\Sbr$ evaluates to $\Sbr \approx 0.105\,\left(\text{V ps}\right)^{-1}$.
The quenching time constant is $\tauQ \approx \ps{9.5}$ for $\Vex = 2$\,V and $\tauQ \approx \ps{6.3}$ for
$\Vex = 3$\,V.
The transit time $d/v^*$ is 5\,ps, i.e.~the adiabatic approximation is valid for excess voltages not
significantly above 3\,V.

\begin{figure}[t!!]
  \centering
  \subfloat[]{\includegraphics[width=7cm]{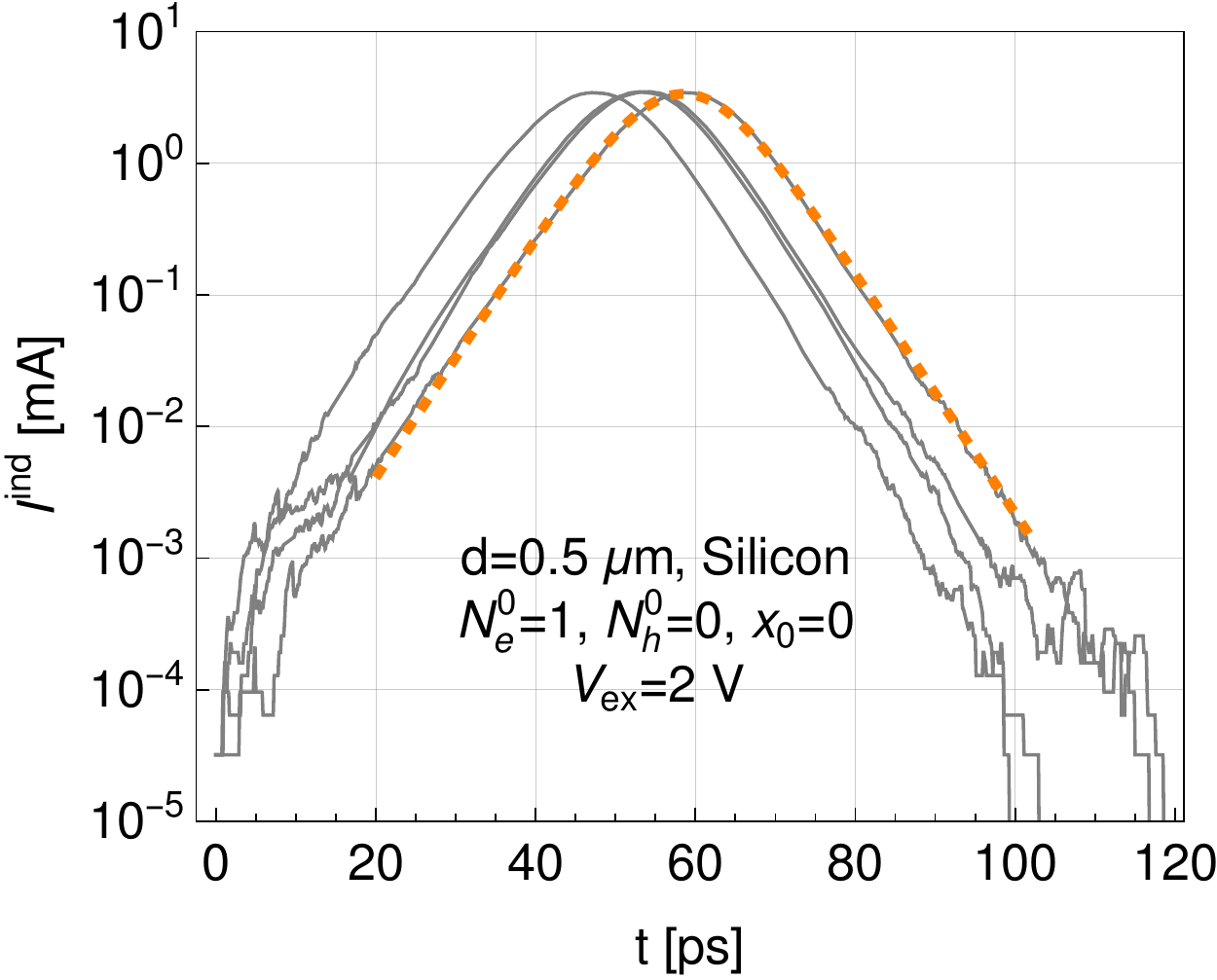}\label{subfig:avalanche_sample_2V}}\qquad
  \subfloat[]{\includegraphics[width=7cm]{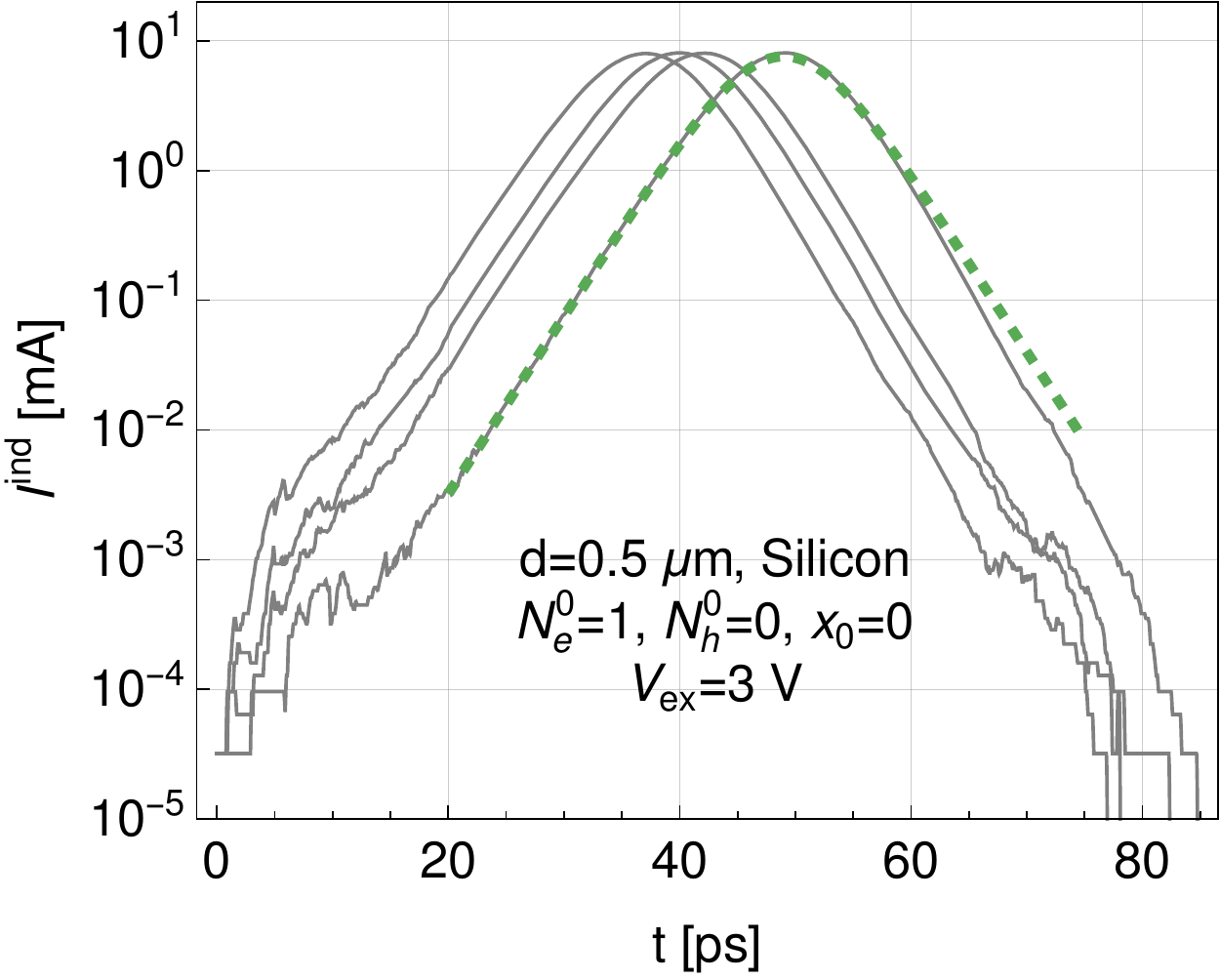}\label{subfig:avalanche_sample_3V}}\qquad
  \caption{(Colour online.) The grey lines represent the induced current $\Iind$ for several avalanche events simulated with the MAMC + circuit model
  for the simplified silicon device described in the main text.
  Each avalanche is initiated by a single primary electron placed at $x_0 = 0$.
  The dashed lines compare the analytic expression for $\Iind$ from Eq.~\ref{eq:Iind_adiabatic_solution_approx} to one avalanche event, 
  for $\Vex = 2$\,V in \protect\subref{subfig:avalanche_sample_2V} and $\Vex = 3$\,V in \protect\subref{subfig:avalanche_sample_3V}.
  }
  \label{fig:example_quenching}
\end{figure}

\begin{figure}[t!!]
  \centering
  \subfloat[]{\includegraphics[height=0.4\textwidth]{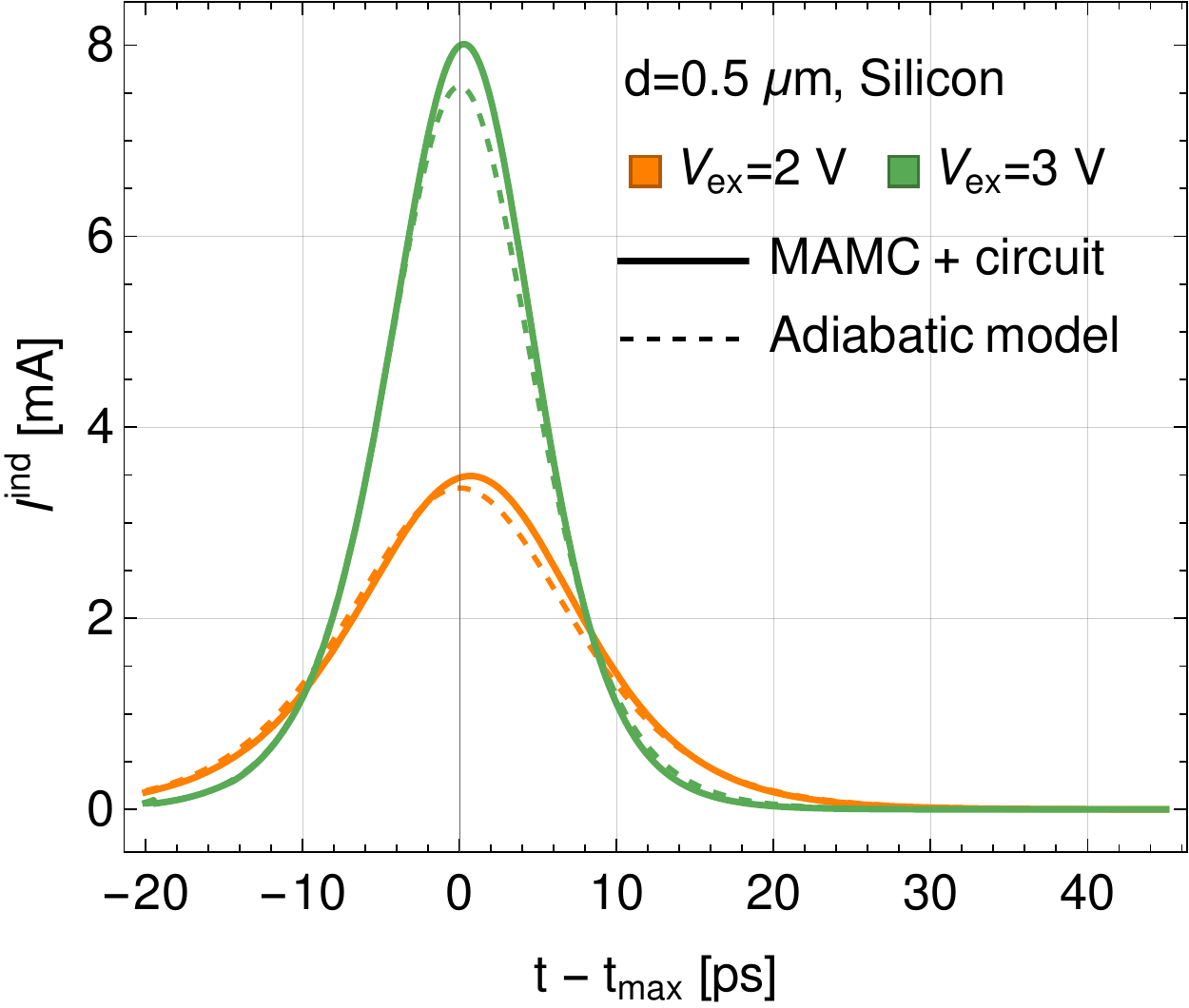}\label{subfig:Iind_without_space_charge}}\quad
  \subfloat[]{\includegraphics[height=0.4\textwidth]{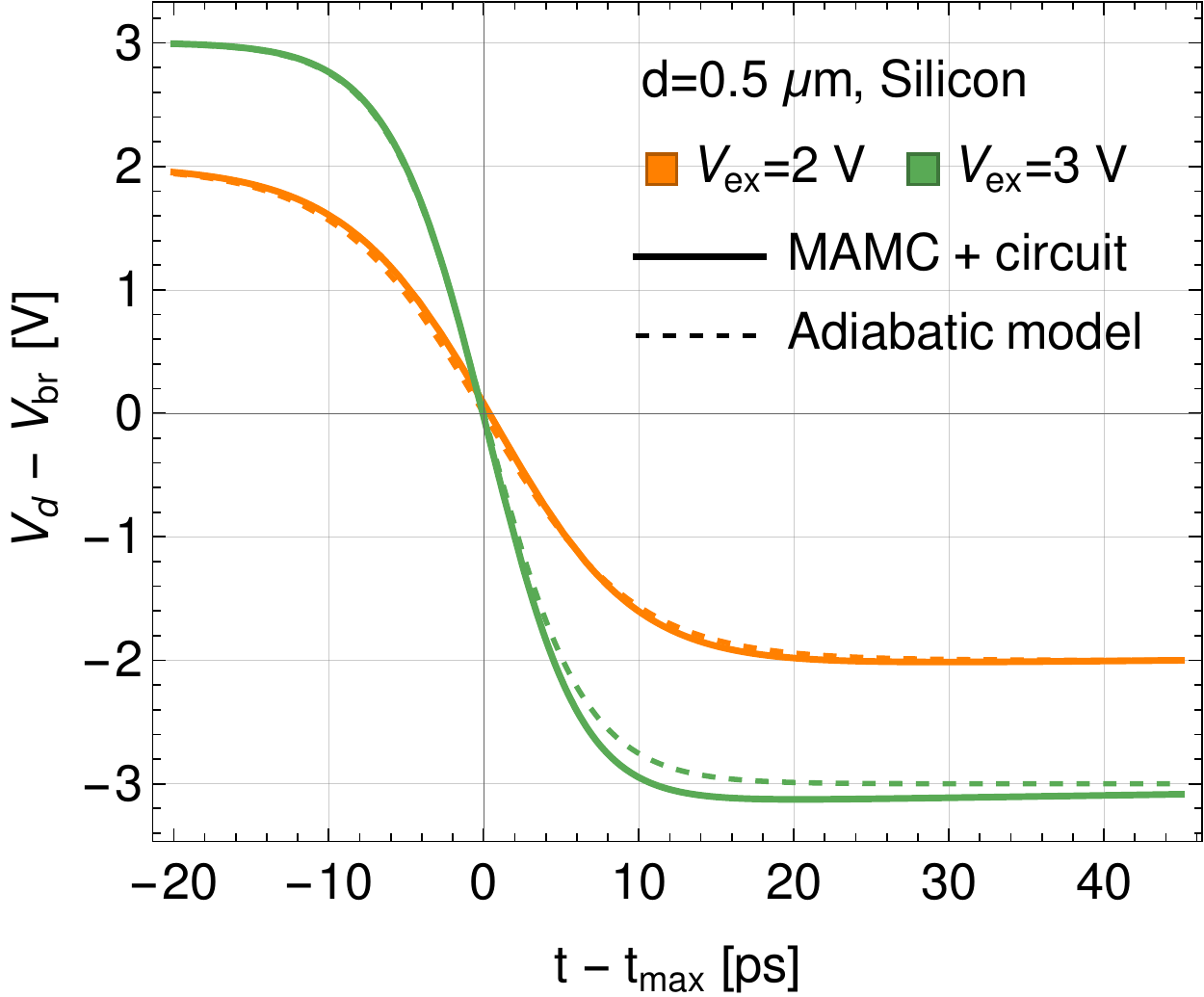}\label{subfig:Vd_without_space_charge}}
  \caption{(Colour online.)
    Comparison between the MC simulation (thick lines) and the adiabatic approximation
    in Eqs.~\ref{eq:adiabatic_solution} (dashed lines), for excess voltages of $\Vex=2$\,V (orange) and $\Vex=3$\,V (green).
    The induced current $\Iind$ is shown in \protect\subref{subfig:Iind_without_space_charge} and the voltage
    $V_d - \Vbr$ in \protect\subref{subfig:Vd_without_space_charge}.
  }
  \label{fig:signal_without_space_charge}
\end{figure}

Fig.~\ref{fig:example_quenching} shows the induced current $\Iind$ for several avalanche events and different excess voltages, obtained
with the MAMC + circuit simulation model.
Fluctuations in the early development of the avalanche lead to a stochastic displacement of the deterministic current waveform from
Eq.~\ref{eq:Iind_adiabatic_solution_approx}.
The simulated time of maximal current $\avg{\tmax}$ is $\ps{55.9}$ ($\ps{40.8}$) for an excess voltage $\Vex$ of $2$\,V ($3$\,V).
These values are well reproduced by Eq.~\ref{eq:tmax_memoryless}, which evaluates to $\avg{\tmax} \approx \ps{56.1}$ ($\ps{40.9}$)
for $\Vex = 2$\,V ($3$\,V).

\begin{figure}[t]
  \centering
  \subfloat[]{\includegraphics[width=0.24\textwidth]{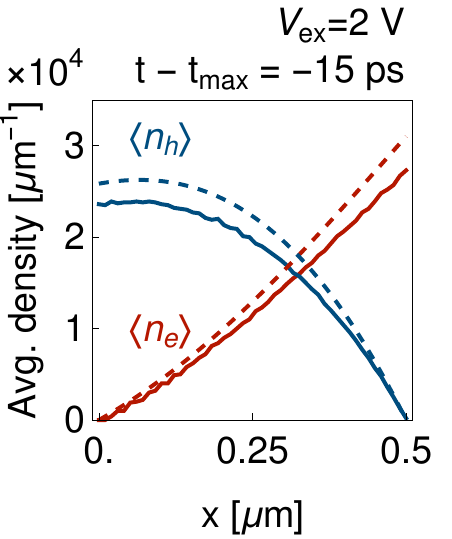}\label{subfig:evolution_wo_space_charge_Vex_2_first}}
  \subfloat[]{\includegraphics[width=0.24\textwidth]{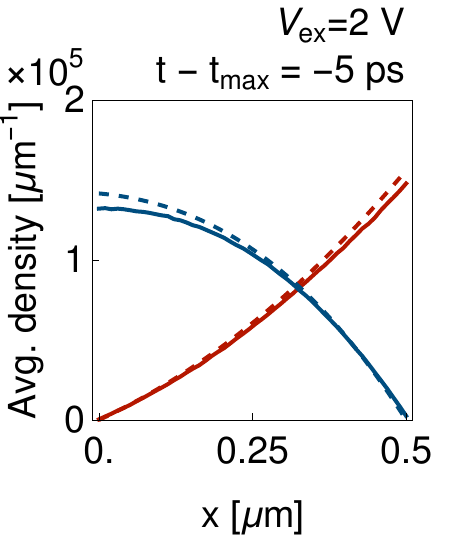}}
  \subfloat[]{\includegraphics[width=0.24\textwidth]{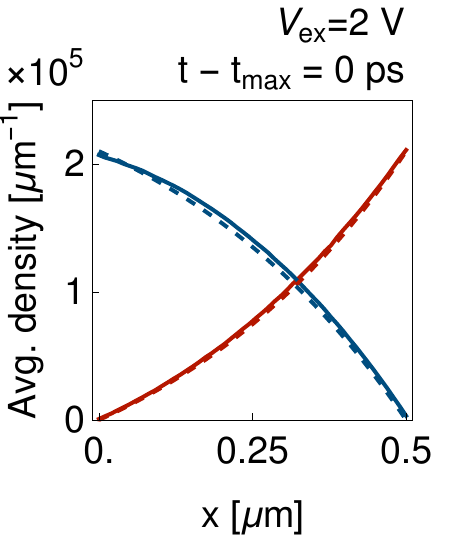}}
  \subfloat[]{\includegraphics[width=0.24\textwidth]{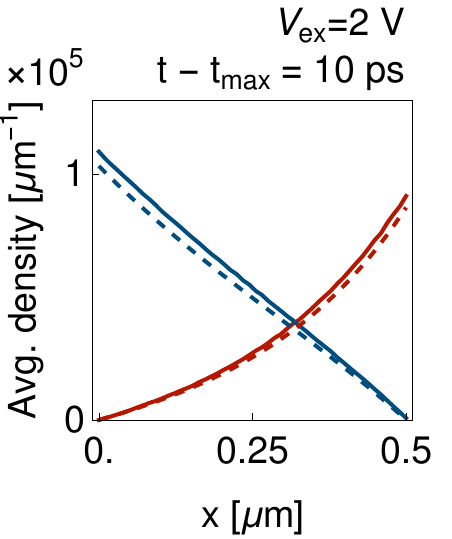}\label{subfig:evolution_wo_space_charge_Vex_2_last}}\\
  \subfloat[]{\includegraphics[width=0.24\textwidth]{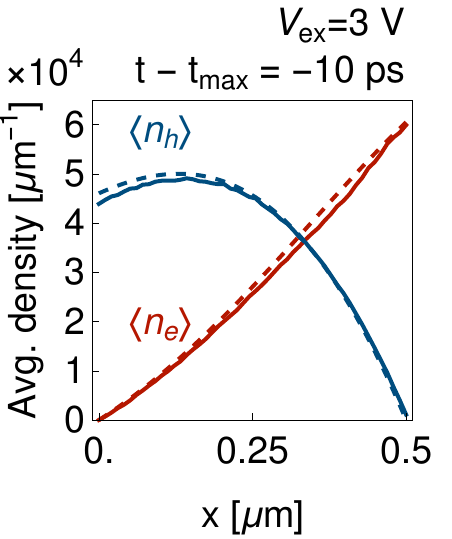}\label{subfig:evolution_wo_space_charge_Vex_3_first}}
  \subfloat[]{\includegraphics[width=0.24\textwidth]{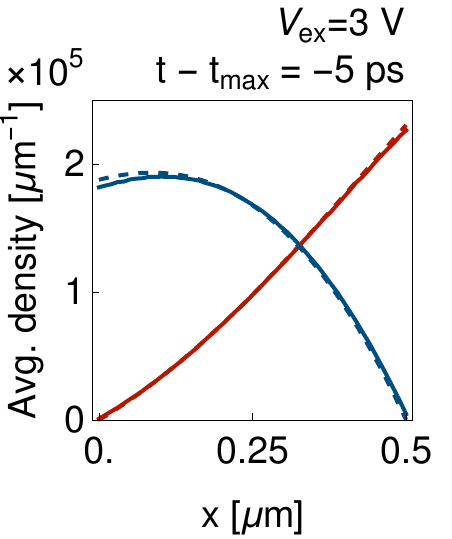}}
  \subfloat[]{\includegraphics[width=0.24\textwidth]{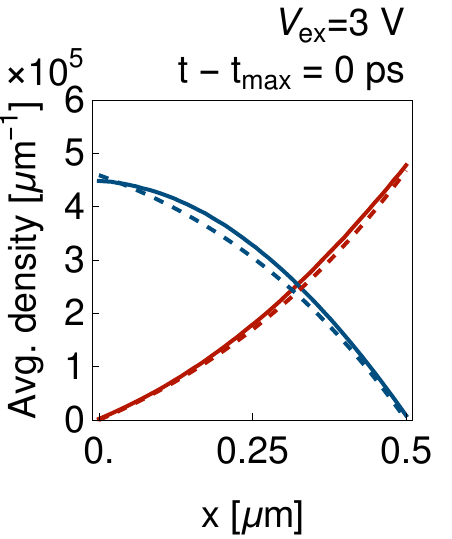}}
  \subfloat[]{\includegraphics[width=0.24\textwidth]{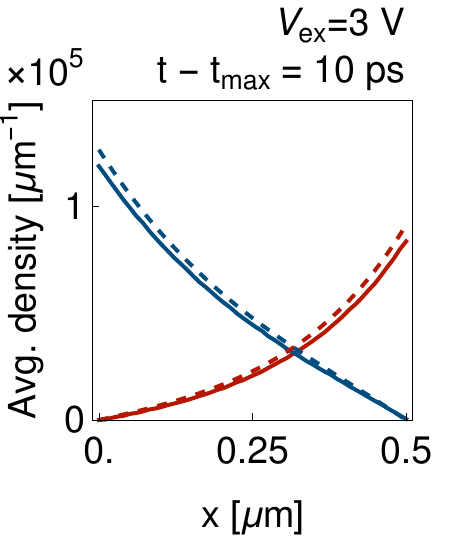}\label{subfig:evolution_wo_space_charge_Vex_3_last}}
  \caption{Evolution of the average charge carrier densities with time, for $\Vex=2$\,V
    in \protect\subref{subfig:evolution_wo_space_charge_Vex_2_first}--\protect\subref{subfig:evolution_wo_space_charge_Vex_2_last}
    and $\Vex=3$\,V
    in \protect\subref{subfig:evolution_wo_space_charge_Vex_3_first}--\protect\subref{subfig:evolution_wo_space_charge_Vex_3_last}.
    The thick lines show the densities obtained from the MAMC + circuit simulation, and the dashed lines correspond to the predictions from the
    adiabatic model.
    All parameters are identical to those used in Fig.~\ref{fig:signal_without_space_charge}.}
  \label{fig:density_comparison_without_space_charge}
\end{figure}

Fig.~\ref{fig:signal_without_space_charge} compares the shapes of the approximate solutions in
Eqs.~\ref{eq:Iind_adiabatic_solution_approx}--\ref{eq:Vd_adiabatic_solution_approx} to the result from the MAMC + circuit simulation
model around the time $\tmax$, demonstrating very good agreement between the analytic result and the simulation.
Small deviations arise as $V_d$ crosses the breakdown voltage, resulting in a slight underestimation of $\Delta V$ for large
excess voltages.
This is caused by additional dynamic effects originating from changes in the shapes of the spatial carrier densities 
of electrons $n_e(x,t)$ and holes $n_h(x,t)$ that are not included in the adiabatic model.
The temporal evolution of these densities is compared in Fig.~\ref{fig:density_comparison_without_space_charge}.
(For the simulation, the average densities $\avg{n_e}$ and $\avg{n_h}$ are shown to ensure a meaningful comparison with the
adiabatic model even at early times where the number of charge carriers is small.)
Charge carriers of both polarities are present throughout the entire multiplication region.
The highest densities are attained at the corresponding downstream ends, i.e.~at $x=d$ for electrons and at $x=0$ for holes.
The shape of the distributions, and in particular the curvature of $n_h(x)$, depends on $V_d$.

\paragraph{Discussion and comparison with other quenching models}

The relation $\Delta V_d = G\,\Vex$ with $G = 2$ is derived in Eq.~\ref{eq:delta_Vd} as an exact analytic result that holds in 
situations where the linear expansion in Eq.~\ref{eq:linear_approximation_S1} is appropriate.

\begin{figure}[t]
    \centering
    \includegraphics[width=0.34\textwidth]{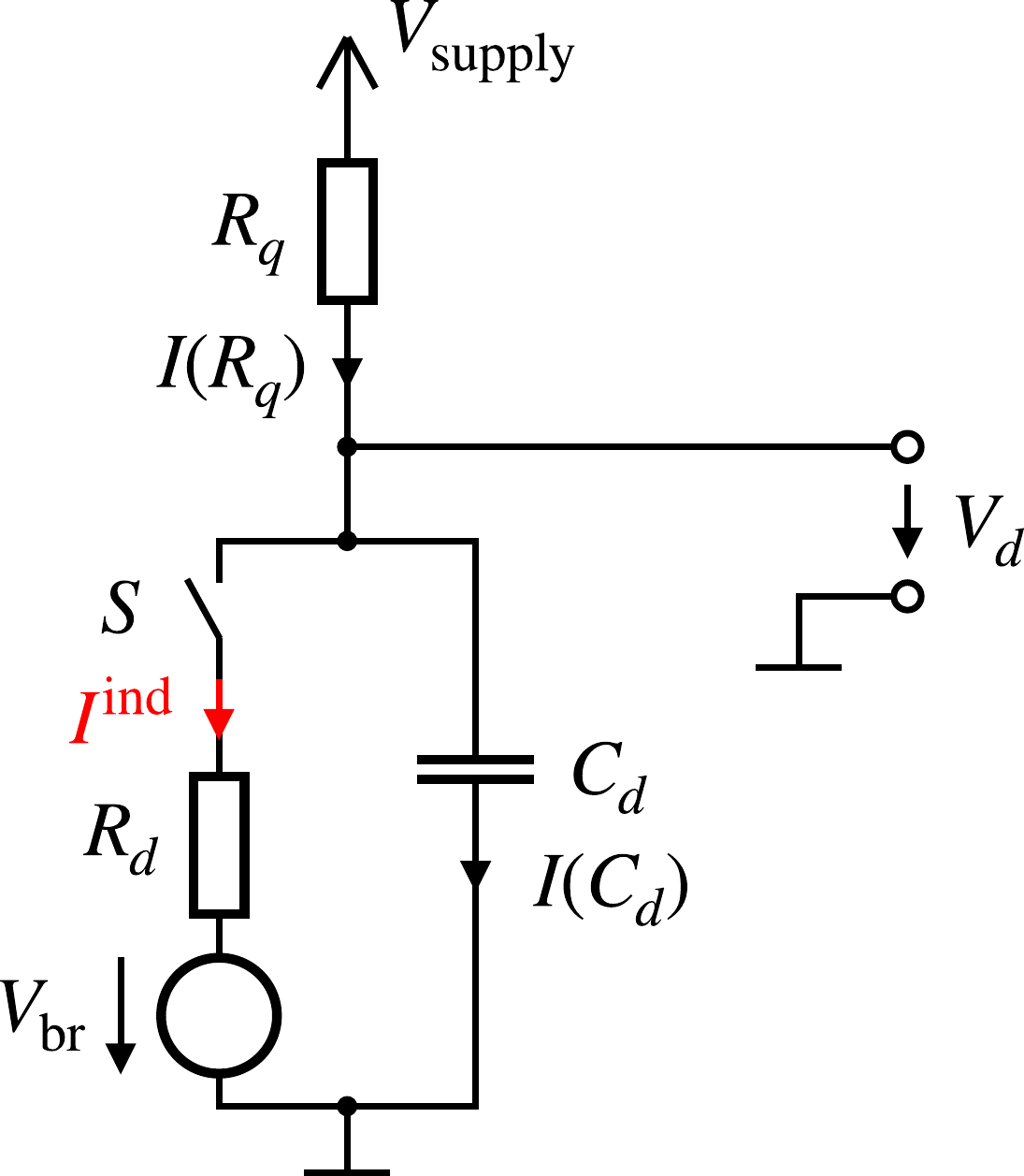}
    \caption{Equivalent circuit used in Refs.~\cite{seifert:2009, cova:1996, haitz:1964} to describe a passively quenched SPAD,
        consisting of the quench resistor $R_q$, the diode capacitance $C_d$ and the dynamic diode resistance $R_d$.
        The switch $S$ controls the state of the avalanche.
        }
    \label{fig:quenching_model_circuit}
\end{figure}

This conclusion differs from predictions made with simpler, widely-used, quenching models \cite{seifert:2009, cova:1996, haitz:1964}.
In these models, the avalanche is taken to directly transition the device from its initial inactive configuration ($V_d = \Vsupply$ and $\Iind = 0$)
into a quasi-time-independent, active state where $V_d = \Vbr$ and a persistent current $I_p = \Vex / R_q$ flows through the quench resistor.
If $R_q$ is chosen sufficiently large, the current $I_p$ is small enough so that avalanche fluctuations may spontaneously quench
the discharge after a brief period of time (the probability per unit time for this to happen is exponentially suppressed in $I_p$ \cite{haitz:1964}).
This leads to a voltage step of $\Delta V_d = \Vex$, i.e.~$G=1$.
The central proposition is that this behaviour can be modelled by the equivalent circuit shown in Fig.~\ref{fig:quenching_model_circuit}, where the 
diode resistance $R_d$ is explicitly included and the avalanche is represented by a switch $S$ whose state marks the presence (closed) or absence (open)
of an avalanche discharge.

The microscopic analysis conducted here indeed confirms the existence of this active equilibrium state:
Eq.~\ref{eq:Iind_evolution} implies $d\Iind/dt = 0$ at $V_d = \Vbr$, which reveals $\Iind = I(R_q) = I_p$ as a valid configuration of 
the circuit in Fig.~\ref{subfig:quenching_circuit}.
This shows that the simple model in Fig.~\ref{fig:quenching_model_circuit} can adequately describe the possible active (switch closed) and inactive
(switch open) equilibrium configurations of the device.

There is, however, no a-priori reason to expect this passive circuit to correctly model the fast transition between these two states.
In particular, the avalanche current $\Iind$ in Fig.~\ref{fig:quenching_model_circuit} decays exponentially as $V_d$ approaches
$\Vbr$, and $V_d \geq \Vbr$ throughout the entire quenching process.
This is apparently inconsistent with the definition of $\Vbr$ as the voltage above which each existing carrier produces a diverging
number of secondary charges \cite{haitz:1964}.
To the best of our knowledge, the existing literature does not address (or resolve) this contradiction for switch-based models.

The microscopic treatment reveals that it is not the absolute value of the avalanche current (or, equivalently, the total number 
of charges in the avalanche) that vanishes close to the breakdown point, but rather its growth rate $S_1$, thus
giving rise to a different transient behaviour and a proportionality factor $G = 2$.
A similar factor $G \approx 2$ has also been obtained in Refs.~\cite{inoue:2020, inoue:2021} through the numerical solution of a system of 
rate equations describing the avalanche, and in Ref.~\cite{hayat:2010} through a microscopic MC simulation similar to the one presented 
here.
Refs.~\cite{inoue:2020, inoue:2021} furthermore report measurements of $\Delta V_d$ and find $G > 1$.

\section{Space charge effects} 
\label{sec:space_charge}
\noindent
For sufficiently large avalanches, the space charge field $\vEsc(\vx, t)$ caused by the charge carriers in the avalanche
may become comparable to the field $\vEext(x, V_d(t))$ produced as a result of the externally applied voltage $V_d$.
The electric field profile $\vE$ that is relevant for the development of the avalanche is now the sum of both components,
i.e.~$\vE = \vEext + \vEsc$.
A full description of the effects of space charge (SC) on the quenching process requires a three-dimensional simulation of the avalanche together with the computation
of a self-consistent solution for the space charge field at every time step.
In particular, $\vEsc$ depends on the spread of the avalanche in the direction transverse to the externally impressed field $\vEext$.

\begin{figure}[tp]
  \centering
  \includegraphics[width=0.45\textwidth]{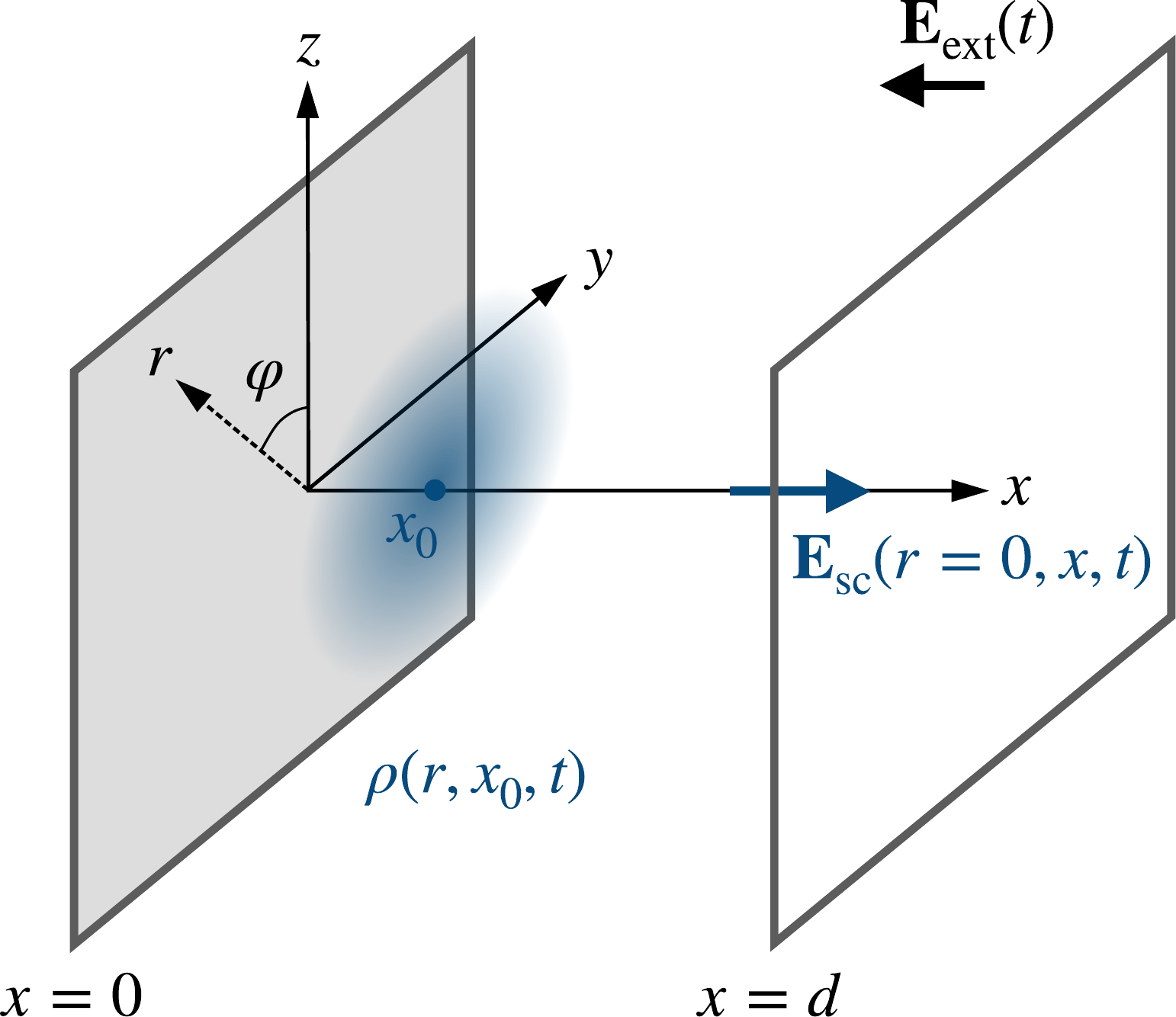}
  \caption{
    For the computation of the space charge field, the high-field region is taken to be delimited by two conducting plates of
    infinite transverse extent, located at $x=0$ and $x=d$.
    Charge carriers are constrained to move along the $x$-axis. This charge distribution is extended into a three-dimensional
    charge density $\rho(r, x, t)$, illustrated here in the plane $x=x_0$.
    The field $\vEsc(r, x, t)$ is evaluated on the $x$-axis as the electric field generated by $\rho(r, x, t)$.
    The external field $\vEext(t)$ also contributes to the total electric field $\vE(x, t)$ relevant for the development
    of the avalanche.
  }
  \label{fig:space_charge_simulation_setup}
\end{figure}

The situation is studied here in a simplified way as illustrated in Fig.~\ref{fig:space_charge_simulation_setup}.
The evolution of the avalanche continues to be simulated by the MAMC + circuit simulation model, with impact ionisation
resulting from the drift of charge carriers along the $x$-axis.
A parameterised model is used to describe the development of the avalanche in the transverse $yz$-plane, defining a radially symmetric
charge profile $\rho(r, x, t)$.
Different physical effects may be included in this parameterisation, e.g.~charge carrier diffusion as well as drift due to
nonvanishing radial electric field components.
The prescribed charge distribution $\rho$ is then used to compute the space charge field $\vEsc$ along the $x$-axis,
which, by symmetry, only has a nonvanishing $x$-component, $\vEsc(r=0,x,t) = \Esc(x, t) \, \ex$.
This (one-dimensional) field distribution is used to determine the impact ionisation coefficients $\alpha(x,t)$ and $\beta(x,t)$.
The finite transverse extent of the avalanche is thus taken into account when computing its space charge field, but
it is \textsl{neglected} when evaluating the backaction of this field on the further evolution of the avalanche. 
This is analogous to the ``1.5-dimensional'' model of space charge effects in resistive plate chambers (RPCs) discussed in
Ref.~\cite{rpc}.
As before, the field $\vEext$ is assumed to be position-independent, i.e.~$\vEext = -\ex \, V_d / d$.
The resulting simulation model is referred to as ``MAMC + circuit + SC''.

For the results presented here, only transverse diffusion is considered for the construction of $\rho(r,x,t)$.
This simplified model is likely to overestimate the true effect of space charge on the development of the avalanche.
First, the space charge field $\vEsc$ itself is overestimated: the $x$-component of the field originating from the chosen
radially symmetric charge density $\rho(r,x,t)$ attains a maximum at $r=0$ (cf.~Eqs.~\ref{eq:rho_e} and \ref{eq:rho_h} below).
All drifting charge carriers are thus subject to the maximal field, while regions of lower field strengths at $r>0$ are not probed
by the avalanche.
Second, the transverse size of the avalanche is underestimated by only considering diffusion, but ignoring carrier drift.
This leads to a narrower distribution of charges and also enhances the computed space charge field.

\paragraph{Transverse spread of the avalanche and space charge field}

Charge carrier diffusion is modelled by prescribing a normal distribution for the transverse distribution of charge carriers, while
the longitudinal distribution continues to be given by the densities $n_e(x, t)$ and $n_h(x, t)$.
The densities $\rho_e$ and $\rho_h$ that are relevant for the computation of the space charge field are thus
\beqarr
\rho_e(r, x, t) &=& -\frac{e_0 n_e(x, t)}{2\pi \sigma_e^2(t)} \exp\left[-\frac{r^2}{2\sigma_e^2(t)}\right],\label{eq:rho_e}\\
\rho_h(r, x, t) &=& \frac{e_0 n_h(x, t)}{2\pi \sigma_h^2(t)} \exp\left[-\frac{r^2}{2\sigma_h^2(t)}\right].\label{eq:rho_h}
\eeqarr
The transverse extent is governed by the transverse diffusion coefficients $D_T$ for electrons and holes,
i.e.~$\sigma_e(t) = \sqrt{2 D_{T,e} t}$ and $\sigma_h(t) = \sqrt{2 D_{T,h} t}$.
Values of $D_{T,e} = 15\,\mathrm{cm}^2/\mathrm{s}$ \cite{bartelink:1970} and
$D_{T,h} = 11\,\mathrm{cm}^2/\mathrm{s}$ \cite{hinckley:1995} are used in the following.
To the best of our knowledge, no measurements of the transverse diffusion coefficients are available in the literature for
the very strong longitudinal fields $|\vEext|>\Vum{20}$ that occur above breakdown.
The selected values correspond to a field of $\Vum{5}$ in silicon.
The transverse diffusion coefficients depend only relatively mildly on the longitudinal electric field,
and it is expected that these values give at least an approximate description of diffusion in the relevant regime.
This is sufficient for the qualitative assessment of space charge effects attempted here.

The electrostatic Green's function for the parallel-plate geometry of Fig.~\ref{fig:space_charge_simulation_setup}
derived in Refs.~\cite{heubrandtner:2002, riegler:2016} is used to compute the space charge field $\vEsc$.

\paragraph{Quenching dynamics with space charge}

\begin{figure}[t]
  \centering
  \subfloat[]{\includegraphics[width=0.24\textwidth]{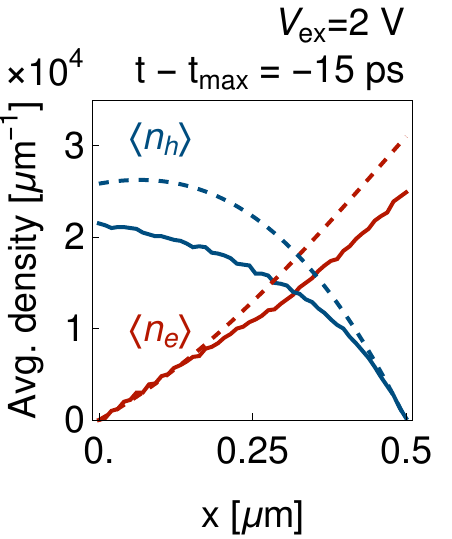}\label{subfig:density_w_space_charge_first}}
  \subfloat[]{\includegraphics[width=0.24\textwidth]{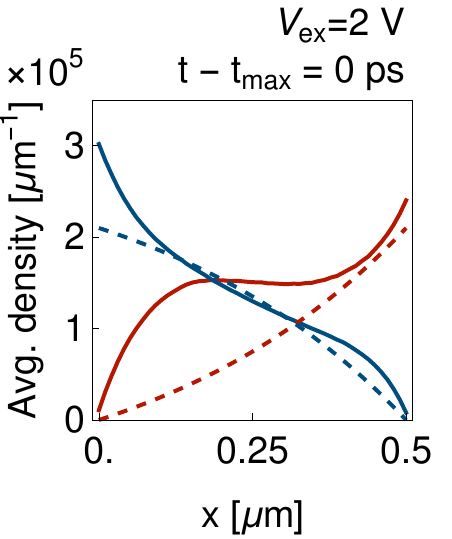}}
  \subfloat[]{\includegraphics[width=0.24\textwidth]{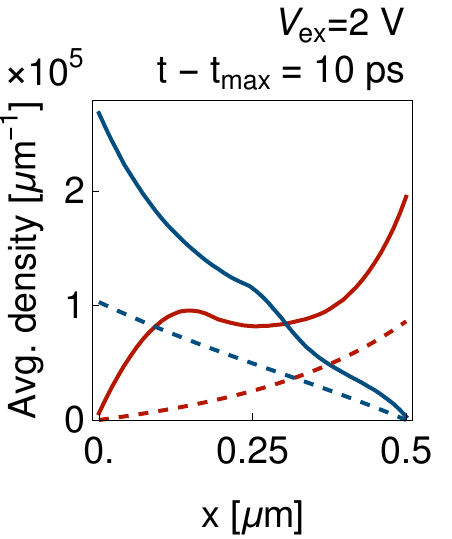}\label{subfig:density_w_space_charge_last}}
  \subfloat[]{\includegraphics[width=0.24\textwidth]{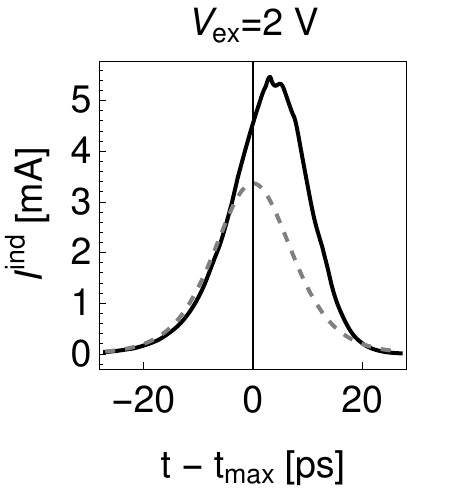}\label{subfig:Iind_w_space_charge}}\\
  \subfloat[]{\includegraphics[width=0.24\textwidth]{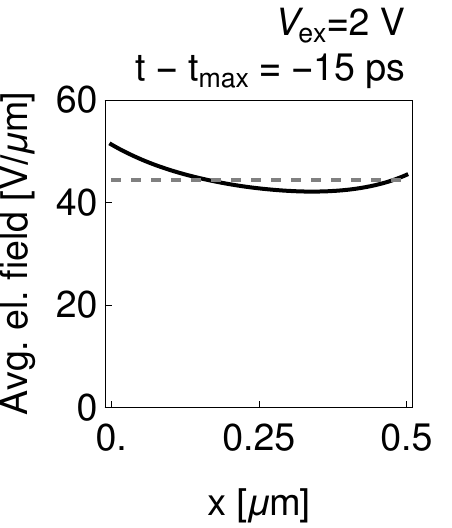}\label{subfig:field_w_space_charge_first}}
  \subfloat[]{\includegraphics[width=0.24\textwidth]{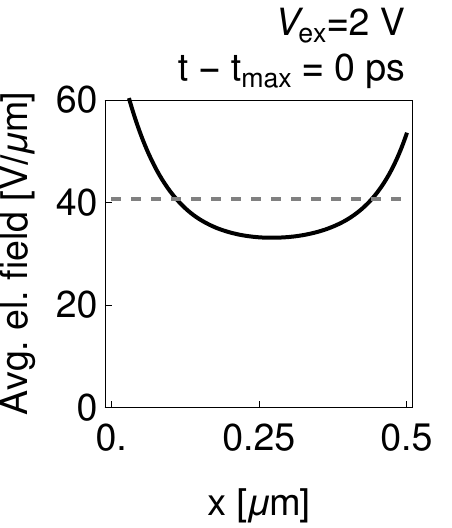}}
  \subfloat[]{\includegraphics[width=0.24\textwidth]{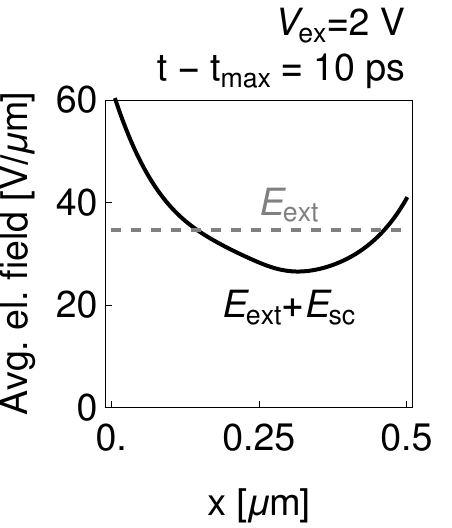}\label{subfig:field_w_space_charge_last}}
  \subfloat[]{\includegraphics[width=0.24\textwidth]{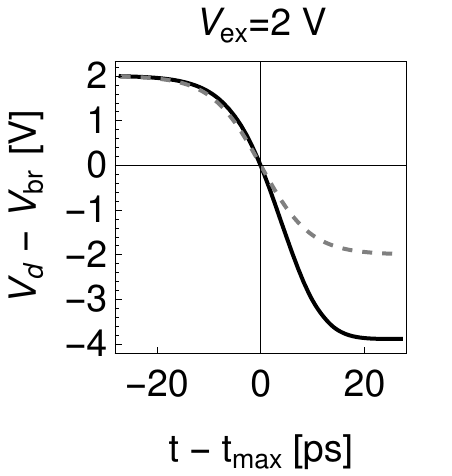}\label{subfig:Vd_w_space_charge}}
  \caption{
    \protect\subref{subfig:density_w_space_charge_first}--\protect\subref{subfig:density_w_space_charge_last} Evolution of the
    average charge carrier densities with time, when space charge effects are included, for $\Vex = 2$\,V.
    The electric field configurations corresponding to these time steps are shown in
    \protect\subref{subfig:field_w_space_charge_first}--\protect\subref{subfig:field_w_space_charge_last}.
    The induced current $\Iind$ is shown in \protect\subref{subfig:Iind_w_space_charge} and the voltage
    $V_d$ in \protect\subref{subfig:Vd_w_space_charge}.
    In all plots, the thick lines correspond to results from the MAMC + circuit + SC simulation, and dashed lines 
    denote predictions from the adiabatic model.
    All parameters are identical to those used in Fig.~\ref{fig:signal_without_space_charge}.
  }
  \label{fig:density_comparison_with_space_charge}
\end{figure}

Fig.~\ref{fig:density_comparison_with_space_charge} shows the evolution of the carrier densities, the electric field, and the
induced signal.
The same simulation parameters and material properties as in Section \ref{sec:passive_quenching} are used.
At early times, space charge effects are of subleading importance, and the resulting charge carrier densities are described by the adiabatic
model in good approximation.
The separation of charge caused by $\vEext$ leads to a space charge field which reduces the electric field in the centre of the gain
layer and enhances it close to its boundaries.
(Space charge cannot alter the voltage $V_d$ measured across the junction contacts, i.e. $\int_0^d d\vs \cdot \vEsc = 0$.)
This results in increased impact ionisation rates and carrier densities near the edges of the multiplication region.
For the chosen material parameters and device geometry, this effect overcompensates the suppressed avalanche growth in regions with
reduced field.
The signal charge and the voltage step $\Delta V_d$ increase by about 50\% compared to the situation without space charge effects,
but the signal shape is not significantly altered.

This is very different from the space charge effects in RPCs studied in detail in Ref.~\cite{rpc}.
Also in the RPC geometry the field integral throughout the sensitive volume remains unchanged so there will be regions with reduced 
field and regions with increased field. 
Since in a gas detector there is however only electron multiplication and no ion multiplication, the electrons end up only in these 
low field regions and the gain is suppressed by many orders of magnitude.

Measurements of the voltage step $\Delta V_d$ in SPADs can thus serve as a direct probe of the importance of space charge effects
for the avalanche development.

\section{Conclusions}
\noindent
In a passively quenched SPAD, the diverging avalanche discharges the detector capacitance, which in turn
lowers the electric field in the multiplication region and reduces the probability for impact ionisation.
The dynamics of this process is here described quantitatively based on a microscopic model of the avalanche 
development, applicable as long as space charge effects are negligible.
The main conclusions of this treatment are:
\begin{itemize}
    \item The duration of the avalanche discharge scales with the quenching time constant $\tauQ$ (cf.~Eq.~\ref{eq:tauQ_def_approx}).
    It is determined in terms of the excess voltage $\Vex$ and the parameter $\Sbr$ (cf.~Eq.~\ref{eq:linear_approximation_S1}).
    The latter may be extracted starting from the field profile $\vE(x)$ around the breakdown point.
    It is known analytically for position-independent fields (cf.~Eq.~\ref{eq:model_parameter}).
    \item The avalanche attains its maximum size as the voltage $V_d$ measured across the diode crosses the breakdown voltage.
    The total voltage step $\Delta V_d$ caused by the discharge is therefore $\Delta V_d = 2 \Vex$.
    The total signal charge is $Q = 2 \Vex C_d$, with the junction capacitance $C_d$ (cf.~Eqs.~\ref{eq:delta_Vd} and \ref{eq:signal_Q}).
    \item Analytic expressions for the evolution of $V_d(t)$ and the current $\Iind(t)$ induced by the avalanche are available
    in Eqs.~\ref{eq:Iind_adiabatic_solution_approx} and \ref{eq:Vd_adiabatic_solution_approx}.
    Analytic formulae for the spatial charge carrier densities also exist.
\end{itemize}
The space charge field produced by the charge carriers in the avalanche enhances the electric field
(and therefore impact ionisation) close to the boundaries of the multiplication region and reduces it in 
its interior.
Charge carriers are present throughout the entire junction, and the avalanche growth may, in principle,
get enhanced or suppressed by the altered field configuration.
A simplified MC simulation of a representative silicon device geometry shows a modest enhancement
of the signal charge as a result of space charge.
This is in stark contrast to gas detectors such as RPCs, where space charge considerably reduces the gas gain.

Our results in Eqs.~\ref{eq:delta_Vd} and \ref{eq:signal_Q} deviate from the prevalent description of passive quenching, in which 
the avalanche discharge is modelled as a bistable switch.
We encourage further experimental work to investigate and clarify this discrepancy.

\section*{Appendix}
\noindent
To simulate the development of the avalanche for equal carrier drift velocities, $v_e = v_h = v^*$, the avalanche region
$x \in [0, d]$ is partitioned into $N$ equidistant bins of width $\Delta x = d / N$.
The units are chosen such that charges drift a distance of $\Delta x$ during the simulation time step
$\Delta t = \Delta x / v^*$.
The bin centre of bin $j$ is located at coordinate $x^j$ and the simulation time steps are indexed as $t^i = i \Delta t$.

The electric field distribution $\vE(x, t) = -E^x(x, t)\, \ex$ and the resulting impact ionisation coefficients $\alpha(x, t)$
and $\beta(x, t)$ are discretised into $E^j(t^i) = E^x(x^j, t^i)$, $\alpha^j(t^i) = \alpha(x^j, t^i)$, and
$\beta^j(t^i) = \beta(x^j, t^i)$.
The charge carrier densities $n_e(x,t)$ and $n_h(x,t)$ are represented by the number of electrons $N_e^j(t^i)$ and holes $N_h^j(t^i)$
in each bin.
These carriers are assumed to be uniformly distributed within the bin.

\begin{figure}[tp]
  \centering
  \subfloat[]{\raisebox{0.5mm}{\includegraphics[height=0.3\textwidth]{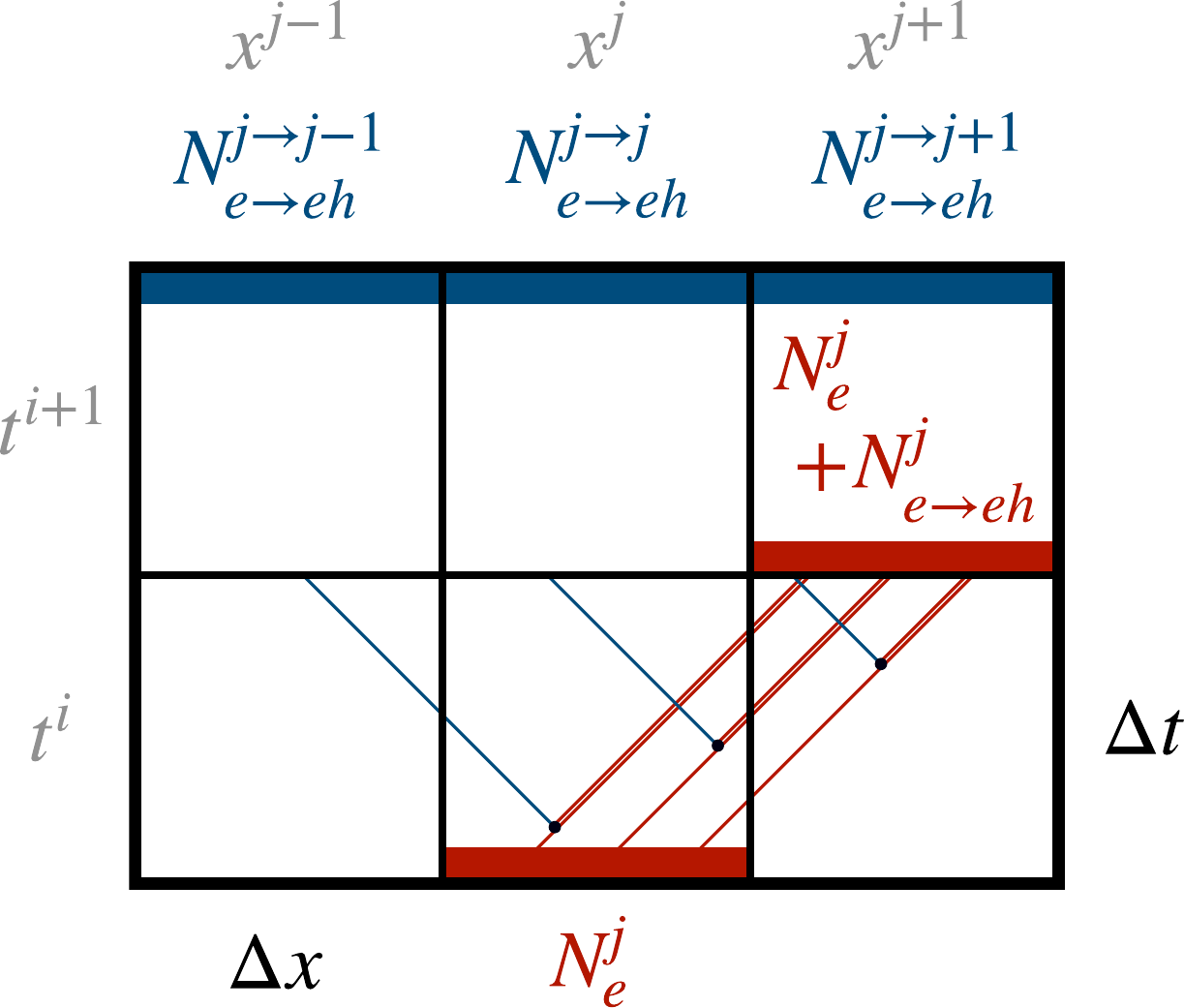}}\label{subfig:e_update_rule}}\qquad\qquad
  \subfloat[]{\includegraphics[height=0.3\textwidth]{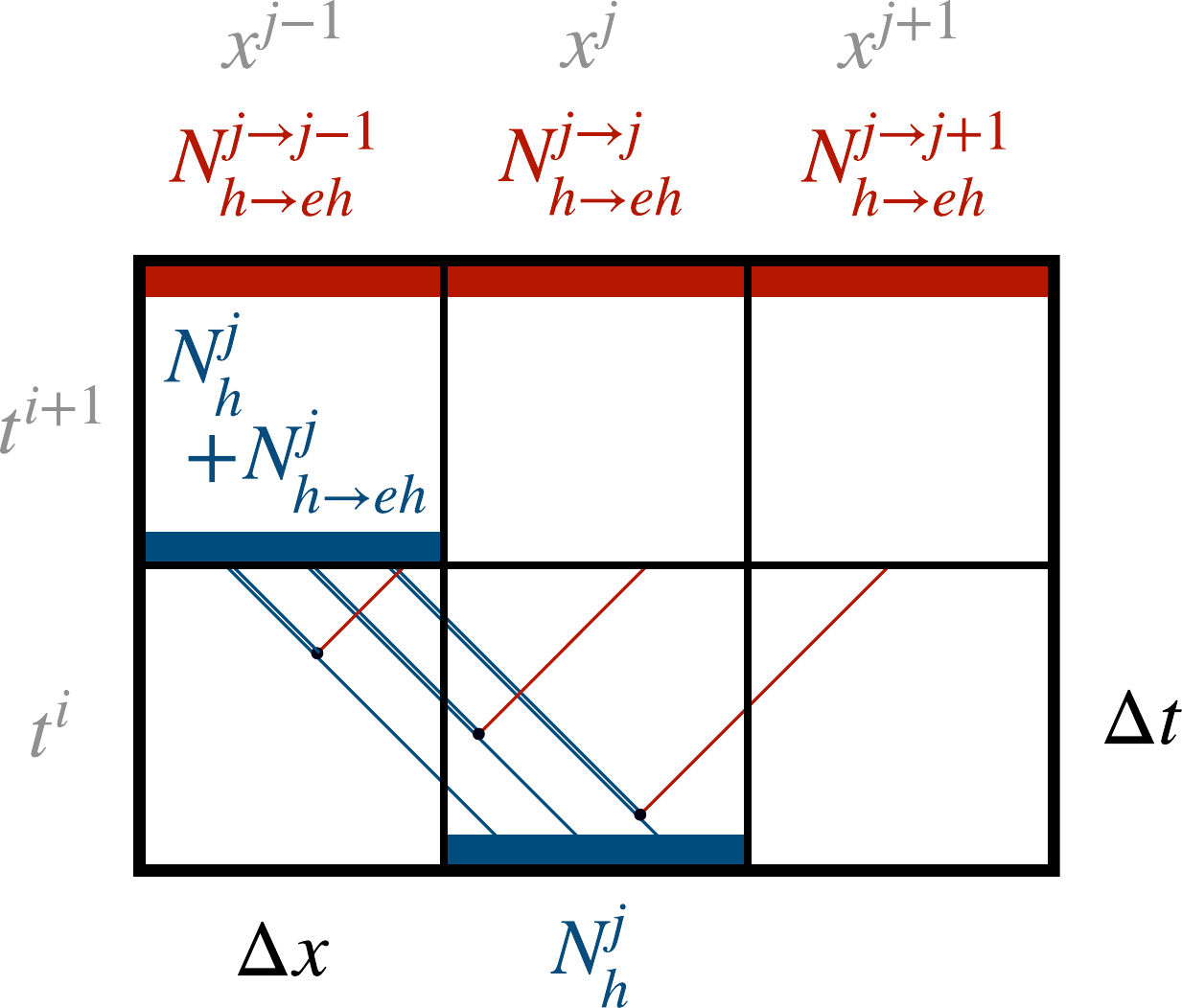}\label{subfig:h_update_rule}}
  \caption{(Colour online.)
    Illustration of the update step $t^i\rightarrow t^{i+1}$ for the evolution of electrons in \protect\subref{subfig:e_update_rule}
    and for the evolution of holes in \protect\subref{subfig:h_update_rule}.
    The number of electrons $N_e^j$ and holes $N_h^j$ contained in a bin $j$ are represented by coloured bars within the respective bin.
    A number of exemplary drift lines are shown to illustrate the movement of charge carriers.
    Impact ionisation events are indicated by black dots.
  }
  \label{fig:MC_update_rule}
\end{figure}

In the absence of impact ionisation, the $\Nej$ holes present in bin $j$ at time $t^i$ drift into bin $j+1$ at $t^{i+1}$, and the
$\Nhj$ holes move into bin $j-1$.
Impact ionisation is simulated as illustrated in Fig.~\ref{fig:MC_update_rule}.
During the time interval $\Delta t$, the $\Nej$ electrons ($\Nhj$ holes) in bin $j$ produce $\Neehj$ ($\Nhehj$) additional
electron-hole pairs.
These numbers are drawn from a Poisson distribution with rate parameters $\alpha^j \Delta x$ and $\beta^j \Delta x$, respectively,
\beqarr*
\Neehj(t^i) &\sim& \Po\left(\Nej(t^i); \alpha^j(t^i)\Delta x\right), \\
\Nhehj(t^i) &\sim& \Po\left(\Nhj(t^i); \beta^j(t^i)\Delta x\right).
\eeqarr*
Impact ionisation occurs uniformly along $\Delta x$.
The $\Neehj$ electrons created in bin $j$ at $t^i$ drift into bin $j+1$ at $t^{i+1}$, while the $\Neehj$ holes spread across
multiple bins: $\NeehjL$ holes reach bin $j-1$, $\NeehjC$ remain in bin $j$, and $\NeehjR$ enter bin $j+1$.
The same consideration holds for the $\Nhehj$ electrons created from hole-initiated impact ionisation events, leading to
$\NhehjL$, $\NhehjC$, and $\NhehjR$ electrons in bins $j-1$, $j$, and $j+1$, respectively.
These numbers are drawn from a multinomial distribution with event probabilities of $p_{j\rightarrow j-1} = 0.25$,
$p_{j\rightarrow j} = 0.5$, and $p_{j\rightarrow j+1} = 0.25$, respectively,
\beqarr*
\{\NeehjL(t^i), \NeehjC(t^i), \NeehjR(t^i)\} &\sim& \Mult\left(\Neehj(t^i); \{p_{j\rightarrow j-1}, p_{j\rightarrow j}, p_{j\rightarrow j+1}\}\right), \\
\{\NhehjL(t^i), \NhehjC(t^i), \NhehjR(t^i)\} &\sim& \Mult\left(\Nhehj(t^i); \{p_{j\rightarrow j-1}, p_{j\rightarrow j}, p_{j\rightarrow j+1}\}\right).
\eeqarr*
The updated charge carrier distributions are then computed as
\beqarr*
\Nej(t^{i+1}) &=& N_e^{j-1}(t^i) + N_{e\rightarrow eh}^{j-1}(t^i) + N_{h\rightarrow eh}^{j+1\rightarrow j}(t^i) + N_{h\rightarrow eh}^{j\rightarrow j}(t^i) + N_{h\rightarrow eh}^{j-1\rightarrow j}(t^i),\\
\Nhj(t^{i+1}) &=& N_h^{j+1}(t^i) + N_{h\rightarrow eh}^{j+1}(t^i) + N_{e\rightarrow eh}^{j+1\rightarrow j}(t^i) + N_{e\rightarrow eh}^{j\rightarrow j}(t^i) + N_{e\rightarrow eh}^{j-1\rightarrow j}(t^i).
\eeqarr*
Assuming a uniform weighting field $E_w / V_w = 1 / d$, the current $\Iind(t^i)$ induced by the drifting charges is computed as
\begin{equation*}
\Iind(t^i) = \frac{e_0 v^*}{d} \sum_j\left(N_e^j(t^i) + N_h^j(t^i)\right),
\end{equation*}
where $e_0$ is the elementary charge.

In the continuum limit where $\alpha\cdot\Delta x  \rightarrow 0$ and $\beta\cdot\Delta x \rightarrow 0$, this simulation model converges to 
the avalanche model described in Section \ref{sec:avalanche_properties}. 
For the results presented in the main body, the avalanche region is divided into $N = 500$ bins.


\end{document}